\documentclass[aip,apl,twoside]{revtex4-1}

\usepackage[tbtags]{amsmath}
\usepackage{amssymb,graphicx}
\usepackage{epstopdf}
\usepackage[alsoload=synchem]{siunitx}
\usepackage[english]{babel}
\usepackage{subcaption}
\usepackage{makecell}
\usepackage{setspace}
\usepackage[justification=raggedright,font={stretch=1.5},labelfont=bf]{caption}
\usepackage{scrextend}
\usepackage[ margin=1.03in, bmargin=.75in]{geometry}

\usepackage{fancyhdr}

\usepackage{bibunits}

\newcommand\SNR{S\!N\!R}
\newcommand\DR{D\!R}
\newcommand\e{\mathrm{e}}
\newcommand\kb{\mathrm{k}_\mathrm{B}}

\newenvironment{supplementary}
    { 	\pagebreak
	\newpage
	\newgeometry{twoside,bottom=2.54cm,top=2.36cm,left=3.23cm,right=3.23cm}

	\setcounter{equation}{0}
	\setcounter{page}{1}
	\setcounter{figure}{0}

	\renewcommand{\thepage}{\arabic{page}}

    }
    { 
     }

\fancypagestyle{fancysupplementary}{

	\fancyhead{}
	\fancyheadoffset{1.6cm}
	\fancyhead[RO]{\Large{\uppercase{Supplementary Information}}}
	\fancyhead[LE]{\Large{\uppercase{Supplementary Information}}}

	\fancyfoot{}
	\fancyfootoffset{1.6cm}
	\fancyfoot[RO]{\textsf{\textbf{\footnotesize{S\thepage}}}}
	\fancyfoot[LE]{\textsf{\textbf{\footnotesize{S\thepage}}}}

	\makeatletter
	\def\p@subsection{}
	\def\p@subsubsection{}
	\makeatother

}

\draft 
\begin{document}

\title{Improving mechanical sensor performance through larger damping} 

\author{Swapan K. Roy}
\affiliation{Department of Physics, University of Alberta}
\affiliation{Nanotechnology Research Centre, National Research Council, Edmonton, Canada}

\author{Vincent T. K. Sauer}
\affiliation{Nanotechnology Research Centre, National Research Council, Edmonton, Canada}
\affiliation{Department of Biological Sciences, University of Alberta}

\author{Jocelyn N. Westwood-Bachman}
\affiliation{Department of Physics, University of Alberta}
\affiliation{Nanotechnology Research Centre, National Research Council, Edmonton, Canada}

\author{Anandram Venkatasubramanian}
\affiliation{Nanotechnology Research Centre, National Research Council, Edmonton, Canada}
\affiliation{Department of Biological Sciences, University of Alberta}

\author{Wayne K. Hiebert}
\email[Corresponding author: ]{wayne.hiebert@nrc-cnrc.gc.ca}

\affiliation{Nanotechnology Research Centre, National Research Council, Edmonton, Canada}
\affiliation{Department of Physics, University of Alberta}

\begin{abstract}

Mechanical resonances are used in a wide variety of devices; from smart phone accelerometers to computer clocks and from wireless communication filters to atomic force microscope sensors.  Frequency stability, a critical performance metric, is generally assumed to be tantamount to resonance quality factor (the inverse of the linewidth and of the damping). Here we show that frequency stability of resonant nanomechanical sensors can generally be made independent of quality factor.  At high bandwidths, we show that quality factor reduction is completely mitigated by increases in signal to noise ratio.  At low bandwidths, strikingly, increased damping leads to better stability and sensor resolution, with improvement proportional to damping.  We confirm the findings by demonstrating temperature resolution of \SI{50}{\micro\kelvin} at \SI{200}{\hertz} bandwidth.  These results open the door for high performance ultrasensitive resonant sensors in gaseous or liquid environments, single cell nanocalorimetry, nanoscale gas chromatography, and atmospheric pressure nanoscale mass spectrometry.

\end{abstract}

\pacs{}

\maketitle 

\begin{bibunit}[naturemag]

Nanoelectromechanical systems (NEMS) are known for extraordinary sensitivity.  Mass sensing has reached single proton level, \cite{Chaste2012, Hiebert2012} enabling NEMS gas chromatography, \cite{Venkatasubramanian2016, Bargatin2012} and mass spectrometry \cite{Hanay2015, Sage2015, Naik2009}.  Force sensing has produced single spin magnetic resonance force microscopy \cite{Degen2009}. Torque resonance magnetometry has been revisioned \cite{Losby2015} with applications in spintronics and magnetic skyrmions.  The mechanical quantum ground state has even become accessible \cite{Teufel2011, Chan2011, OConnell2010}.  The best sensitivities, however, have generally been presumed to require the highest quality factors limiting application to vacuum environments and low temperatures.  A host of new applications could result with ultrasensitivity available in air and liquid: biosensing, security screening, environmental monitoring, and chemical analysis.  As an example, our group aims long-term to combine mass spectrometry and gas chromatography functions into one via NEMS sensing in atmospheric pressure.

\begin{figure}[]
\centering
\includegraphics{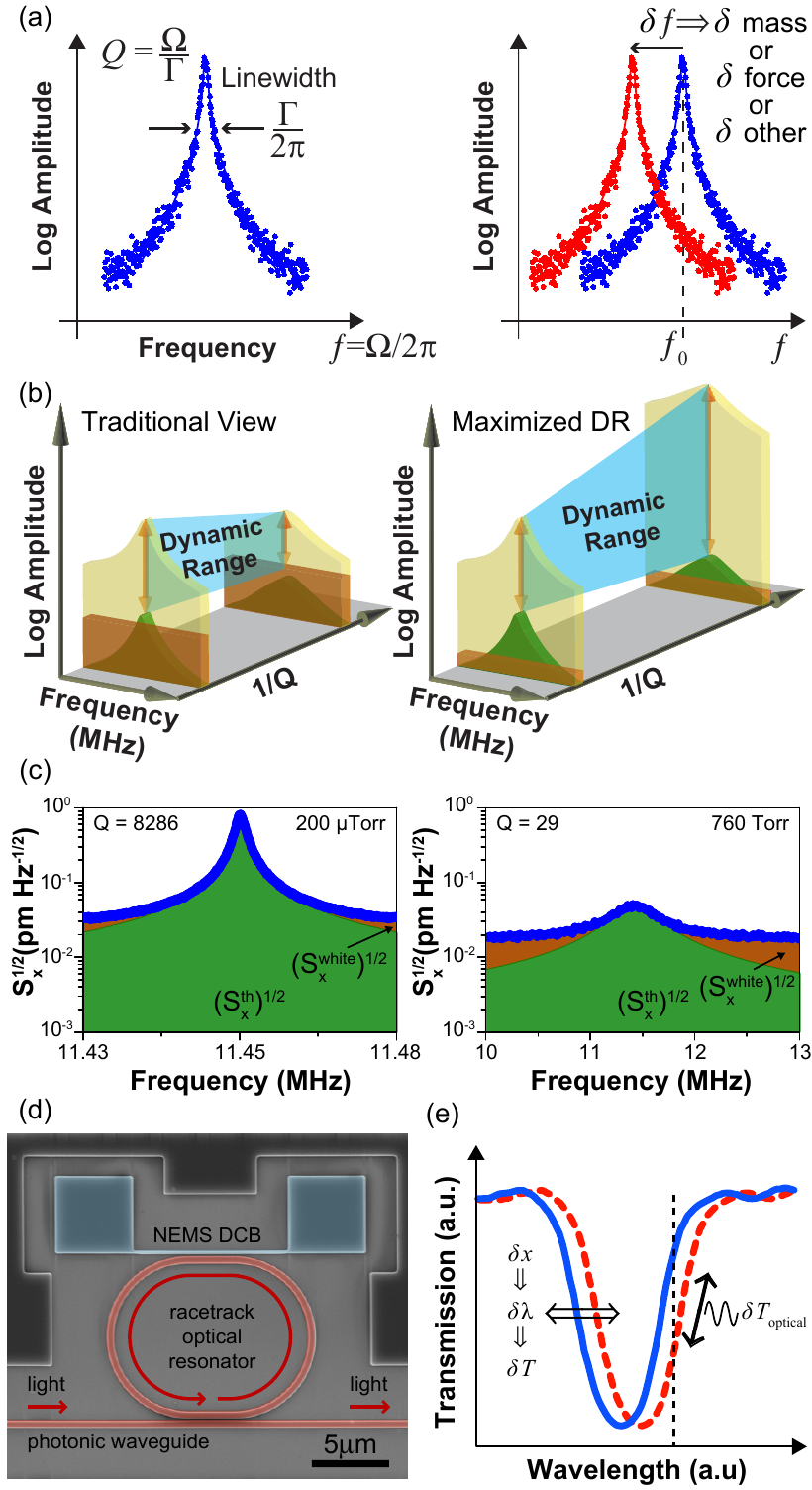}
\caption{(Caption on next page) 
\label{fig:1}}
\end{figure}

Exquisite NEMS sensitivity is enabled through ultra-small mass and stiffness combined with precise resonant frequency determination which allows perturbations to that frequency (such as mass or force) to be probed (see Fig.~\ref{fig:1}a).  Robins' formula \cite{Robins1982}, articulated in the AFM community by Rugar \cite{Albrecht1991} and in NEMS by Roukes \cite{Cleland2002, Ekinci2004}, forms the basis for force and mass sensitivity analyses.  It gives an estimation of the frequency stability based on the resonant quality factor, $Q$, and the comparison of noise energy to motional energy. The formula can be written as follows:
\begin{equation} \label{eq:DRformula}
    \left \langle \frac{\delta f}{f} \right \rangle \sim \frac{1}{2 Q} \frac{1}{\SNR} = \frac{1}{2Q} 10^{-\DR/20},
\end{equation}
where $\SNR$ (signal to noise ratio) is the ratio of driven motional amplitude to equivalent noise amplitude on resonance
\begin{equation} \label{eq:SNR}
    \SNR= \frac{a_{driven}}{a_{noise}} ,
\end{equation}
and the dynamic range $\DR$ is the power level associated with this $\SNR$.  The $Q$ factor in the denominator of equation~\ref{eq:DRformula} has led researchers to pursue high $Q$ for better resolution \cite{Tsaturyan2016, Moser2014, Fong2012}.

However, there is a curious case of when $\SNR \propto 1/Q$ that results in no sensitivity dependence on $Q$.  This is not a special case.  In fact, it is the general case if the $\DR$ is properly maximized.  Conceptually (Fig.~\ref{fig:1}b, right), this follows from duller resonances having a fundamentally lower intrinsic noise floor peak.  At the same time, the wider linewidth tolerates more nonlinearity and extends the linear range to larger amplitude.  Combined, the two effects give $10^{-\DR/20} \propto Q$.

This peculiar observation implies that frequency fluctuation noise should not depend on $Q$ in the case when thermomechanical noise is well resolved and amplitude can be driven to nonlinearity.  No systematic investigation of this startling revelation has been done, even though the model provides a pathway to completely mitigate sensitivity loss due to low $Q$.  This is an exciting prospect with wide-ranging implications for scanning probe microscopy and force sensing, mass sensing and biosensing, and inertial and timing MEMS (gyroscopes/accelerometers and RF oscillators/filters).  Further, a detailed inspection of the phase noise model used in NEMS systems \cite{Albrecht1991, Cleland2002, Ekinci2004} reveals equation~\ref{eq:DRformula} results from an approximation based on long mechanical ringdown times (high $Q$).  Removing this approximation, remarkably, implies frequency fluctuation noise proportional to $Q$ at low bandwidth; thus a highly damped system with full dynamic range should have better frequency stability (and sensitivity) then an equivalent lowly damped one.

Using nano-optomechanical systems, we demonstrate frequency stability \textit{improving} with \textit{increased} damping.  We change pressure from vacuum to atmosphere to vary the extrinsic $Q$ within a single nanomechanical device.  We observe signal to noise ratio growing inversely proportional to $Q$ while the full dynamic range is maintained.  Frequency stability measurements (Allan deviation) within this zone drop with increased damping for a given thermally limited averaging time, approaching closely the theoretical limit.  Notably, the stability at atmospheric pressure is better than that in vacuum.  Also importantly, we see evidence that excess intrinsic frequency fluctuation noise (also known as dephasing/decoherence \cite{Sansa2016, Gavartin2013, Sun2016, Fong2012, Moser2014, Maillet2016}) shrinks with falling $Q$. Intrinsic fluctuation noise does not limit stability at moderate and higher bandwidths, and plays no role at atmospheric pressure.  We go on to test this implied sensitivity improvement with measurements of change in temperature and nanocalorimetry, using the optical ring as calibration, and show \SI{50}{\micro\kelvin} sensitivity at \SI{200}{\hertz} BW.  This is comparable to state of the art \cite{Inomata2016, Zhang2013}, even with the modest calorimeter geometry of a doubly clamped beam, and demonstrates the power of the approach.  These results will allow proliferation of high performance ultrasensitive resonant sensors into gaseous and liquid environments.

\addtocounter{figure}{-1}
\begin{figure} [t!]
  \caption[ Frequency shift sensing, smaller Q can improve dynamic range, and nano-optomechanical system experimental set-up.]{\textbf{Frequency shift sensing, smaller Q can improve dynamic range, and nano-optomechanical system experimental set-up.} (a) Concept of frequency shift sensing: a mechanical resonance is perturbed by change in mass or force resulting in a frequency shift.  At a first approximation, the minimum detectable shift is proportional to sharpness of the resonance, $Q$, and signal to noise ratio, $\SNR$, as per Robins' formula (eq’n~\ref{eq:DRformula}). (b) Concept of maximizing linear dynamic range.  Left: Traditional View.  The dynamic range $\DR$ (arrows) extends from the noise floor to the driven resonance peak (beige Lorentzian-peak shape).  The instrumentation noise floor (brown rectangle) often obscures the thermomechanical noise floor (green Lorentzian-peak shape).  Decreasing $Q$ leads to a loss in system $\DR$.  Right: Maximized $\DR$ case.  Well-resolved thermomechanical noise leads to a drop in noise peak value during increased damping; simultaneously, the upper end of linear range becomes higher as nonlinearity onsets at higher amplitude.  System $\DR$ grows on both ends with falling $Q$. (c) Displacement noise $ S_\mathrm{x}^{1/2}$ (blue circles) of the doubly clamped silicon beam (\SI{9.75}{\micro\meter} x \SI{180}{\nano\meter} x \SI{220}{\nano\meter}) shown in (d).  Left graph is at high $Q$ measured in vacuum; right graph is at low $Q$ measured at atmospheric pressure.  The green fit $ (S_\mathrm{x}^\mathrm{th})^{1/2}$ is resolved out of the orange (white-noise) background $(S_\mathrm{x}^\mathrm{white})^{1/2}$ near resonance.  The peak noise value is suppressed at lower $Q$. (d) Annotated SEM image of the nano-optomechanical system device.  A mechanically released doubly clamped beam (NEMS) is adjacent to a racetrack optical cavity and bus photonic waveguide, all patterned in \SI{220}{\nano\meter} thick silicon-on-insulator.  (e) Concept of the optical cavity resonance shift caused by mechanical beam motion.  Oscillation in displacement $\delta x$ of the mechanical beam modulates the optical resonance wavelength $\delta \lambda$ which, when probe light is situated on the side slope of the cavity, transduces to transmission modulation $\delta T_\mathrm{optical}$. 
}
\end{figure}

\section*{Maximizing dynamic range to minimize frequency fluctuations}

Analyses of ultimate limits for force detection of microcantilevers were carried out early on in the AFM community \cite{Albrecht1991}, narrowing onto thermomechanical noise as the primary limit.  In contrast to macroscale mechanical resonators used as oscillators (such as quartz crystals), the smaller stiffness and size of AFM beams result in non-negligible motion caused from fluctuations of the thermal bath via the equipartition theorem.  In essence, $\frac{1}{2} k_{\textup{B}} T$ of thermal energy populates $\frac{1}{2} k \langle x^2 \rangle$ of modal energy, producing between \SI{}{\pico\meter} and  \SI{}{\nano\meter} average displacements for small stiffness $k$.  These motion levels have been resolvable since the early 1990s.  For mass detection \cite{Cleland2002, Ekinci2004}, reducing mass is paramount, so NEMS-sized devices tend to be stiffer than AFM devices (thermomechanical noise average displacement tends to be in the  \SI{}{\pico\meter} range).  At the same time they are harder to transduce; thus even resolving thermomechanical noise in NEMS had been a challenge in early days \cite{Bunch2007,Li2007}.  With the advent of many new transduction techniques \cite{Unterreithmeier2009, Li2008, Li2007}, thermomechanical noise can now be resolved in NEMS-scale devices on a much more routine basis \cite{Wu2017, Kim2016,Weber2016, Sansa2016, Olcum2015, Gil-Santos2015, Sauer2014, jun2006electrothermal, Gavartin2013, Moser2013, Srinivasan2011, Chan2011, Teufel2011, Gil-Santos2010, Eichenfield2009, Teufel2009}.

Nano-optomechanical systems, in particular \cite{Wu2017, Kim2016, Gil-Santos2015, Sauer2014, Gavartin2013, Srinivasan2011, Chan2011, Teufel2011, Gil-Santos2010, Anetsberger2009, Eichenfield2009, Teufel2009}, have allowed resolving thermomechanical noise by orders of magnitude above the instrumentation noise background.  One example is our microring cavity optomechanical system \cite{Diao2013}, with displacement imprecision of approximately \SI{20}{\femto\meter\per\sqrt\hertz}.  Figure~\ref{fig:1}c shows the measured displacement noise $S_\mathrm{x}^{1/2}$ in an example doubly clamped beam, measured in vacuum where $Q$ is high and at atmospheric pressure where $Q$ is low.  As per convention, values for $S_\mathrm{x}$ are calibrated from voltage signals ($S_\mathrm{V}$) by assuming the peak noise relation (derived via equipartition theorem):
\begin{equation} \label{eq:sxcalibration}
   S_{\textup{x}}^{\textup{th}}\left (f _{\textup{0}} \right )=\frac{4k_{\textup{B}}T}{M\Omega ^{2}\Gamma }
\end{equation}
We define the thermomechancial noise amplitude on resonance $a_\mathrm{th}$ as
\begin{equation} \label{eq:a_th}
   a_\mathrm{th} = \sqrt{S_\mathrm{x}^{\textup{th}}(f_0) \Delta f} = \sqrt{\frac{4k_{\textup{B}}T Q}{M\Omega ^{3}}  \Delta f}
\end{equation}
where $\Delta f$ is the measurement bandwidth.  Details about the thermomechanical noise calibration and displacement imprecision can be found in supplementary information (SI), section 1.2.  In both cases, the noise is dominated by the thermomechanical term near resonance, flattening to a white background far from resonance.  The relatively large peak at high-$Q$ sharply juts out of the background, dominating for \SI{30}{\kilo\hertz}, which is about 20 linewidths.  The suppressed low-$Q$ peak also still reaches out of the background for about 1.5 linewidths (\SI{600}{\kilo\hertz}). It is important to note, equation~\ref{eq:a_th} confirms that $a_\mathrm{th}$ is proportional to $Q^{+1/2}$.  These data show that our system reaches the bottom end of the full dynamic range for at least \SI{30}{\kilo\hertz} measurement bandwidth.

\begin{figure*}[]
\centering
\includegraphics{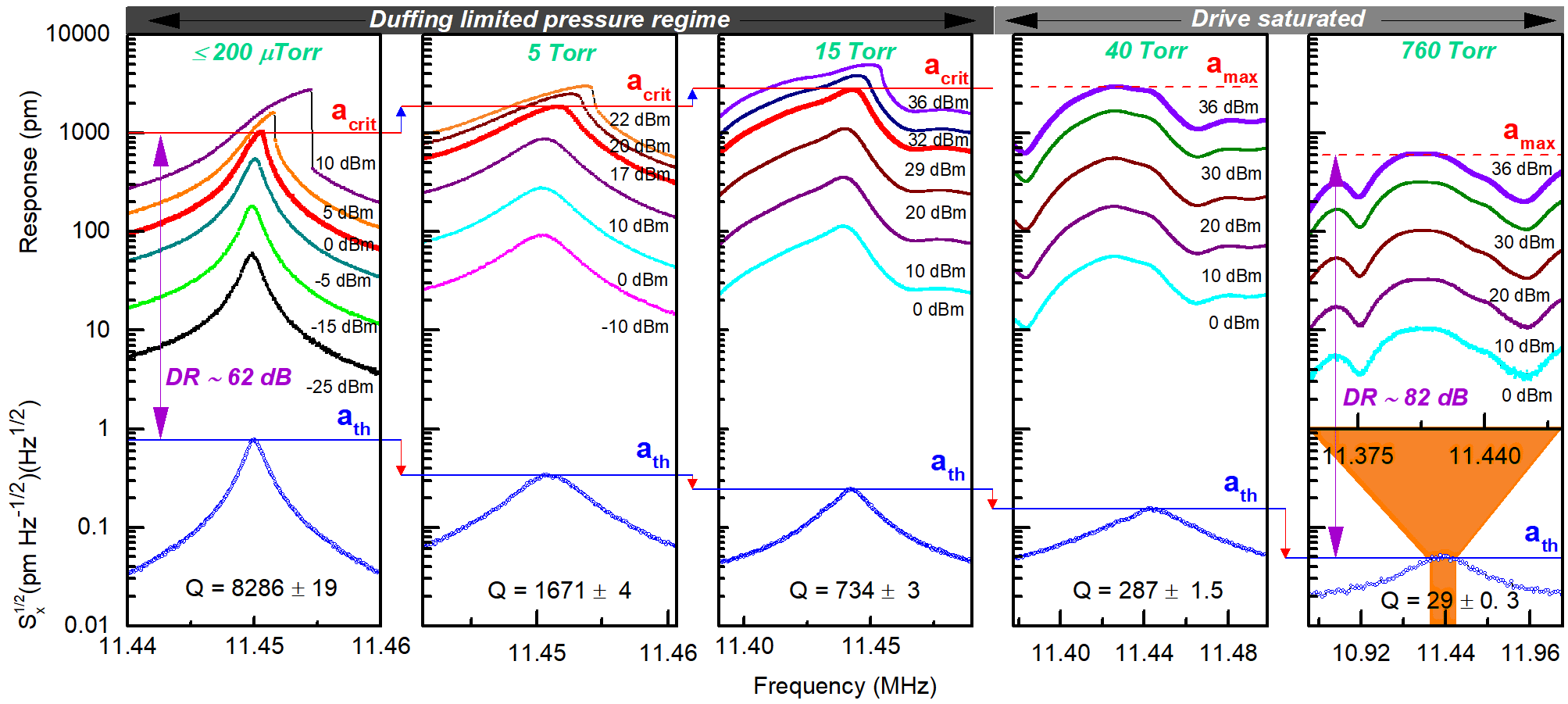}
\caption[Dynamic range is pressure dependent.]{\textbf{Dynamic range is pressure dependent.}  Open symbols (blue) are measured thermomechanical noise frequency curves presented in a \SI{1}{\hertz} bandwidth; $a_\mathrm{th}$ is their peak value which falls with increasing pressure.  Filled symbols are driven response frequency curves for various drive powers; $a_\mathrm{crit}$ (thicker, red) grows with increasing pressure in the Duffing limited pressure regime.  The 760 Torr driven frequency axis is zoomed in with respect to the thermomechanical noise at the same pressure. \label{fig:2}}
\end{figure*}

Our devices are mechanically driven with a shear piezo (see methods) and a large drive power enables the upper end of their linear range to be reached for pressures up to about \SI{30}{\torr}.  As the doubly clamped beam is driven to larger amplitudes, the stiffness becomes amplitude dependent resulting in a geometric nonlinearity \cite{Schmid2016, Kacem2009, Postma2005}.  This Duffing nonlinearity results in sharkfin-shaped resonance traces (Figure~\ref{fig:2} top traces in first 3 panels) and amplitude dependent resonance frequency.  A critical amplitude can be defined to indicate the end of the linear range \cite{Postma2005}:
\begin{equation} \label{eq:acrit}
   a_\mathrm{crit}= \frac{2 (0.745)}{\pi} f_0 L^2 \sqrt \frac{\rho \sqrt 3}{Q E}
\end{equation}
where $L$ is the beam length and $E$ is the Young's modulus (a version of the equation including tension is in the SI). Notice that the critical amplitude is inversely proportional to square root of $Q$ in equation~\ref{eq:acrit}.  The nonlinearity grows and increasingly distorts the lineshape as amplitude grows; naturally, the distortion becomes prominent (i.e. the onset of nonlinearity) at lower amplitude for narrower resonance lines. Taking $a_\mathrm{noise}$ to be $a_\mathrm{th}$ and $a_\mathrm{driven}$ to be $a_\mathrm{crit}$ when the full dynamic range is accessed, equations~\ref{eq:SNR}, ~\ref{eq:a_th}, and~\ref{eq:acrit} combine to produce $\SNR$ proportional to $1/Q$.

In order to test the $\SNR$ behaviour, and its role in equation~\ref{eq:DRformula}, we have measured properties of the same doubly-clamped beam at different pressures (and thus different extrinsic quality factors) from vacuum up to atmospheric pressure.  This approach has the advantage of keeping all parameters except for $Q$ identical.  Results are presented in Figure~\ref{fig:2} with frequency sweeps for five representative pressures.  At each pressure, the thermomechanical noise is plotted for a \SI{1}{\hertz} bandwidth along with the driven root mean square amplitude response for varying drive power.  Marked in thick red are traces for the drive power corresponding with Duffing critical amplitude (up to \SI{15}{\torr}) and in thick purple for the maximum driving power available (40 and \SI{760}{\torr}).  For \SI{15}{\torr} pressures and up, the driven resonance line-shape is distorted.  This is not due to nonlinearity (note the conserved response shape), rather, the resonance has broadened to the point where piezo drive efficiency is no longer a constant function of frequency \cite{Bargatin2008}; the distorted features are related to bulk acoustic resonances in the piezo-chip system.  This distortion carries no information about the nature of the NEMS beam resonance and does not warrant further discussion (See SI, Section 1.5).

The first thing to note in Figure~\ref{fig:2} is that the peak of the noise floor $a_\mathrm{th}$ diminishes as the pressure increases (and $Q$ decreases) and generally follows $a_\mathrm{th} \propto Q^{1/2}$ (cf. eqn~\ref{eq:a_th}).  This can be conceptually understood in the following way.  The area under the thermomechanical resonance curve is conserved for a given temperature (in proportion to $k_{\textup{B}}T$); as the width of the curve increases ($Q$ decreases), the peak value must fall in order to compensate.  For the upper end of the dynamic range, we see that, within the Duffing limited pressure regime, $a_\mathrm{crit}$ is increasing in proportion to $Q^{-1/2}$, as predicted by equation~\ref{eq:acrit}.  Accounting for both effects, $\SNR \propto 1/Q$ up to \SI{15}{\torr} pressure.  At \SI{40}{\torr} and up, we no longer have enough drive power to reach the Duffing critical amplitude and no longer take advantage of the full linear dynamic range of the system.  None-the-less, we note that dynamic range is still higher at atmospheric pressure than it is in vacuum.

\begin{figure}[]
\centering
\includegraphics{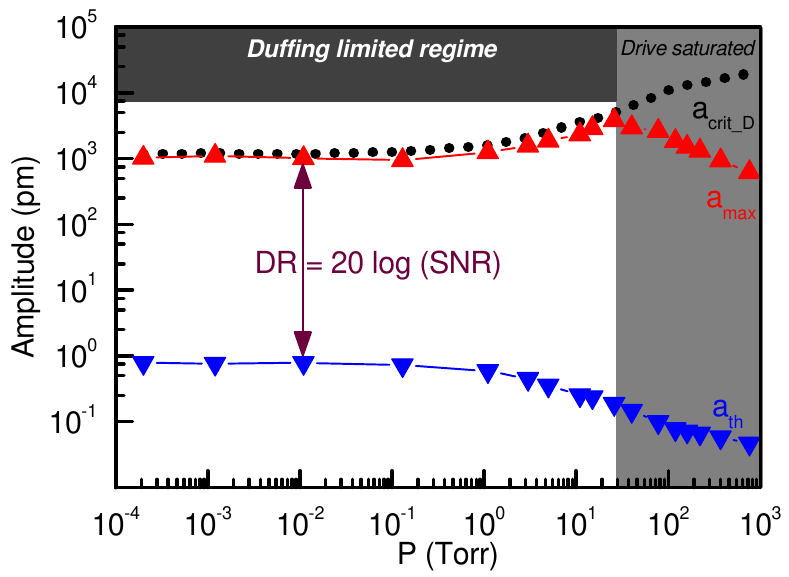}
\vspace{0.5 em}
\includegraphics{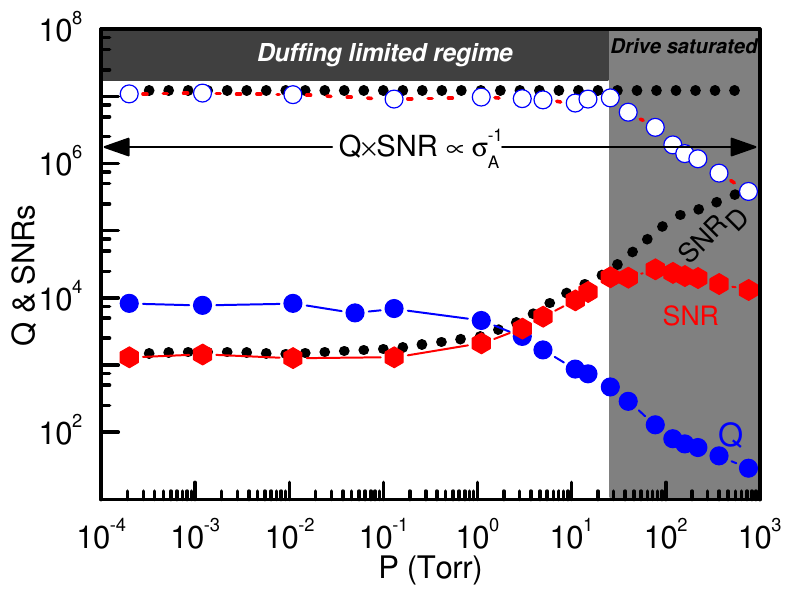}
\caption[The product of $Q\times \SNR$ is constant in the Duffing limited regime.]{\textbf{The product of $Q\times \SNR$ is constant in the Duffing limited regime.} (a) Peak frequency curve amplitude response vs pressure: $a_\mathrm{crit_D}$ is the theoretical Duffing amplitude defined by equation~\ref{eq:acrit}, $a_\mathrm{max}$ is the measured peak amplitude, and $a_\mathrm{th}$ is the thermomechanical peak amplitude.  $DR$ is the dynamic range which grows with pressure.  (b) Quality factor ($Q$) and signal-to-noise ratio ($\SNR$) vs pressure: $\SNR_{\textup{D}}$ is $a_\mathrm{crit_D}/a_\mathrm{th}$ and SNR is $a_\mathrm{max}/a_\mathrm{th}$. \label{fig:3}}
\end{figure}

Figure~\ref{fig:3} plots the peak amplitudes $a_\mathrm{crit}$ and $a_\mathrm{max}$, the thermal amplitude $a_\mathrm{th}$, quality factor $Q$, signal-to-noise ratio $\SNR$, and product of $Q\times \SNR$ as a function of pressure.  From this, we can clearly see that $\SNR$ is inversely proportional to $Q$ and that $Q\times \SNR$ is conserved within the Duffing limited regime.  According to Robins' picture (equation~\ref{eq:DRformula}), the frequency fluctuations in our system should be independent of $Q$ up to \SI{15}{\torr}.

\section*{Frequency fluctuation measurements (Allan deviation)}

With $Q\times \SNR$ conserved, it is left to check the fractional frequency stability $\delta f/f$ in our device.  We do this using the 2-sample Allan variance, a standard method of characterizing frequency stability \cite{Barnes1971} (see SI, section 2.2).  The Allan deviation $\sigma(\tau)$, as the square root of the Allan variance, is an estimate of fractional frequency stability for a given time $\tau$ between frequency readings. The functional form for $\sigma(\tau)$ (subscripted with R to remind of the connection to Robins and Eqn. 1) is

\begin{equation} \label{eq:sigma}
    \sigma_{\mathrm{R}}(\tau) = \frac{1}{4 Q} \frac{1}{\SNR} \frac{1}{\sqrt{\Delta f}} \frac{1}{\sqrt{\tau}} 
\end{equation}

\begin{figure*}[]
\centering
\includegraphics{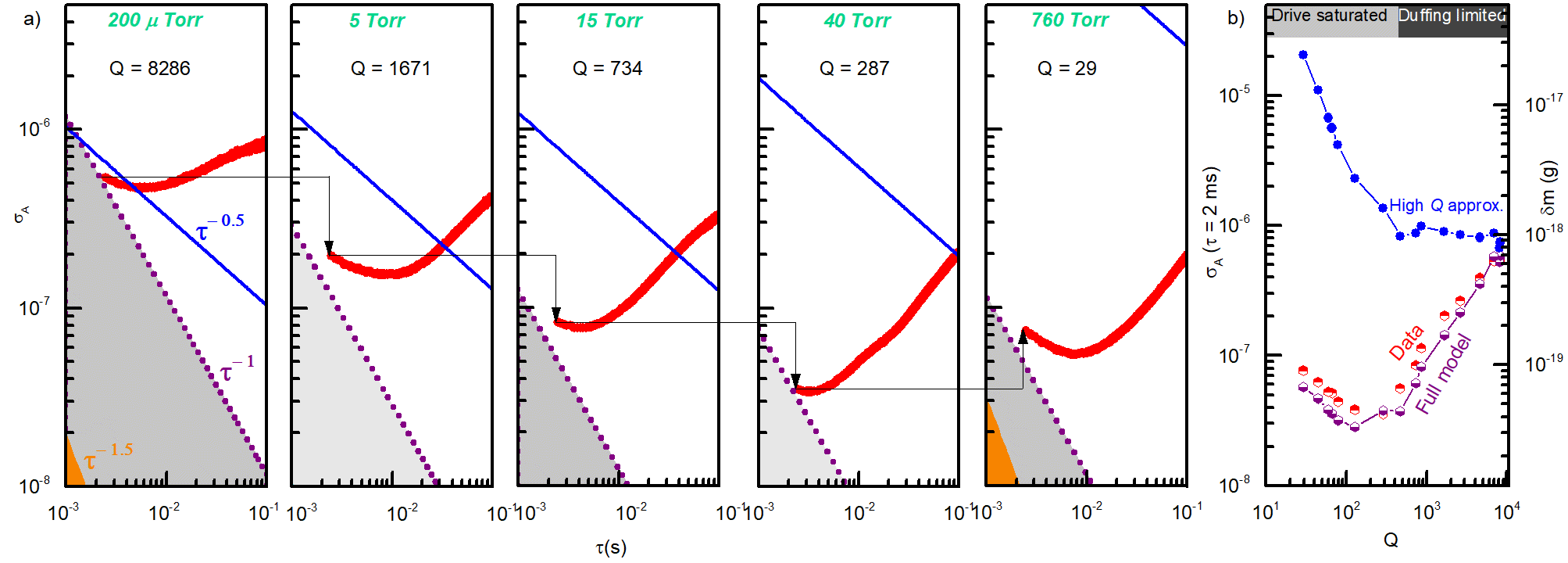}
\caption[Allan deviation $\sigma_ \mathrm{A}$ falls (improves) with falling $Q$.]{\textbf{Allan deviation $\sigma_ \mathrm{A}$ falls  (improves) with falling $Q$.}  (a) Allan deviation (thick red line) vs sampling time at different pressures.  Solid line (blue) is theoretical minimum from equation 1 while dotted line (purple) with shading is theoretical minimum from equation 7.  Shaded (orange) region (only visible in panel 1 and 5) is theoretical minimum set by instrumentation noise floor.  (b) Allan deviation at 2 ms sampling time vs Q.  High Q approximation corresponds to equation 6, full model to equation 10, and data to the experimentally measured values.  In the Duffing limited regime, the data and full model are proportional to Q.  In both regimes, the data reaches close to the fundamental limit of equation 10. \label{fig:5}}
\end{figure*}

Figure~\ref{fig:5} presents the measured Allan deviation data for our device at the 5 representative pressures and $Q$s.  Data is taken with a \SI{4}{\kilo\hertz} demodulation bandwidth and collected while tracking frequency in a \SI{500}{\hertz} phase-locked loop (PLL).  The \SI{4}{\kilo\hertz} represents the integration bandwidth for the noise, while the \SI{500}{\hertz} sets the bound above which the PLL begins to attenuate fluctuations (effectively setting a minimum meaningful $\tau$ for $\sigma(\tau)$).  Details of the Zurich lock-in amplifier and PLL settings can be found in SI, section 1.6.

Astonishingly, rather than staying constant, the Allan deviation is actually improving as the pressure increases and $Q$ falls, up to 40 Torr pressure.  Further, the measured data dip well below the theoretical minimum set by Robins' formalism and equation~\ref{eq:sigma} (solid blue lines).

\section*{Full analysis of Allan deviation from noise power}

To solve this mystery, we need to understand the close connection between Allan deviation and phase noise\cite{Barnes1971}.  The Allan variance is essentially an integration of close-in phase noise $S_\phi (\omega)$, with an appropriate transfer function $H(\tau,\omega)$.  Here, $\omega = 2 \pi f_\mathrm{mod}$, where $f_\mathrm{mod}$ is the frequency-offset-from-carrier $f_\mathrm{mod} = f - f_0$ and the integration goes from zero up to the measurement bandwidth $\Delta f$.  The resulting Allan deviation $\sigma$ will be proportional to $\left \langle S_\phi\times \Delta f \right \rangle ^{1/2}$, where the $\langle \rangle$ brackets here loosely represent the integration.

Understanding the frequency stability then reduces to understanding the behaviour of $S_{\phi}$.  We can define $S_{\phi}^\mathrm{x}$ as the portion of phase noise caused by displacement noise $S_\mathrm{x}$ (full details are available in SI section 2)
\begin{equation} \label{eq:sphix}
S_{\phi}^\mathrm{x} = \frac{1}{2} \frac{S_\mathrm{x}}{a_\mathrm{driven}^2}.
\end{equation}
Close to resonance, the Lorentzian-shaped thermomechanical noise peak (\textit{cf.} Fig. 1c) turns into a low-pass filter with $1/f^2$ rolloff (see Fig. 5b)

\begin{equation} \label{eq:sxomega}
S_\mathrm{x} (\omega) = S_\mathrm{x}(0) \frac{(\Gamma/2)^2}{\omega^2 + (\Gamma/2)^2}.
\end{equation}
Combining equations~\ref{eq:SNR}, \ref{eq:a_th}, \ref{eq:sphix}, and \ref{eq:sxomega} gives $S_{\phi}^\mathrm{x}(\omega)$
\begin{equation} \label{eq:sphixomega}
S_{\phi}^\mathrm{x}(\omega) = \frac{1}{\SNR^2} \left ( \frac{1}{2\Delta f} \right ) \frac{(\Gamma/2)^2}{\omega^2 +  (\Gamma/2)^2}.
\end{equation}

So far, the analysis follows closely to previous Robins' analyses \cite{Albrecht1991, Cleland2002, Ekinci2004}.  At this point, the assumption is generally made that $\omega^2 + (\Gamma/2)^2 \approx \omega^2$, \textit{i.e.} that $Q$ is high.  This assumption turns Eqns.~\ref{eq:sxomega} and \ref{eq:sphixomega} from low pass filters into pure rollofs (see Fig. 5a).  In particular, knowing that $S_\mathrm{x}(0) \propto 1/\Gamma$ (\textit {cf.} eqn. ~\ref{eq:sxcalibration}), it is concluded that $S_{\phi}^\mathrm{x} \sim S_{x} \sim \Gamma^{+1}$, and ultimately that $\sigma \propto \Gamma^{+1/2}$.  This is a generally well-known result in the AFM community.

\begin{figure*}[]
\centering
\includegraphics{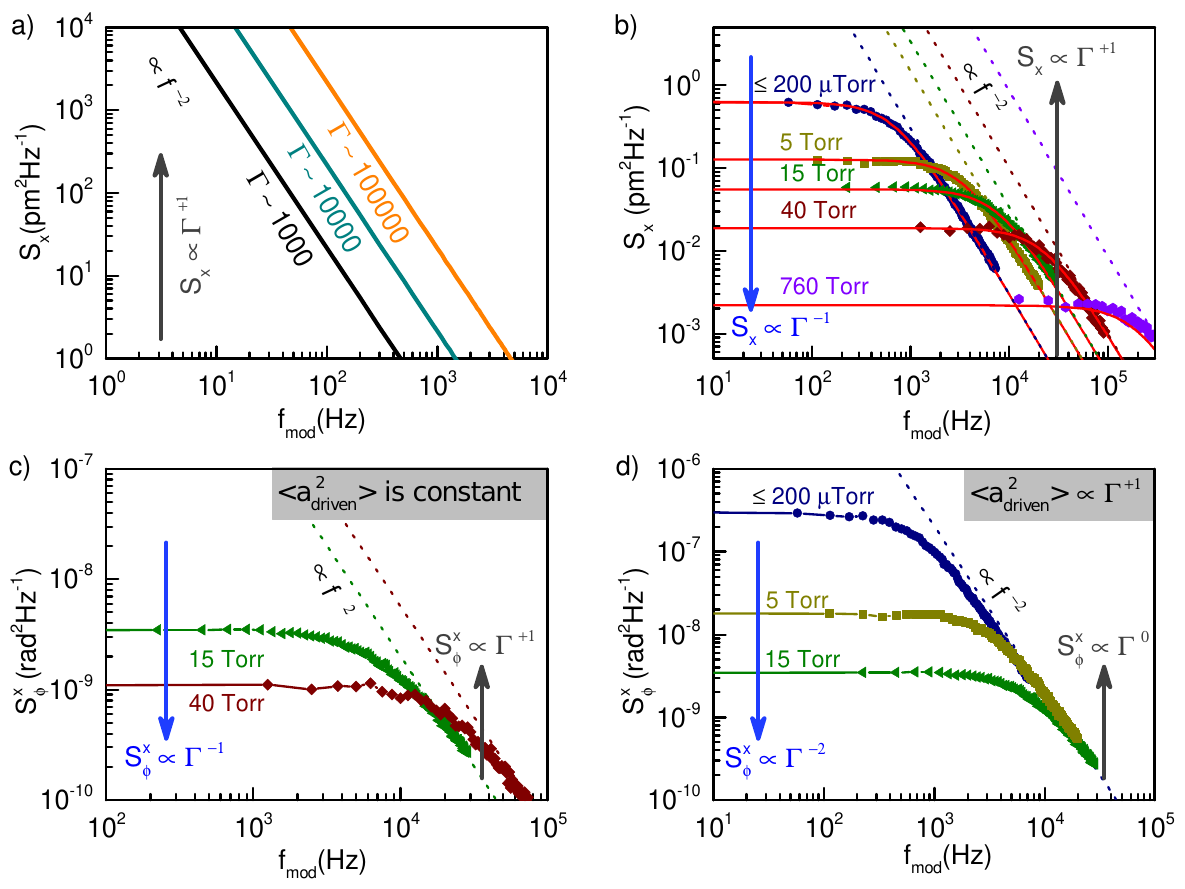}
\caption[Noise power behaviour with respect to damping can be proportional, constant, inversely proportional, and inversely quadratic.]{\textbf{Noise power behaviour with respect to damping can be proportional, constant, inversely proportional, and inversely quadratic.}  (a) Concept of thermomechanical displacement noise being proportional to damping for pure rolloff.  (b) Measured thermomechanical noise fit to equation 5; noise is proportional to damping above the rolloff, inversely proportional below.  (c) Measured displacement noise converted to phase noise with constant driven amplitude; noise is proportional to damping above the rolloff and inversely proportional below.  (d) Measured displacement noise converted to phase noise with squared driven amplitude proportional to damping; noise is independent of damping above the rolloff and inversely quadratic below. \label{fig:4}}
\end{figure*}

Something interesting happens when the high $Q$ assumption is not made.  Figure 5b shows our experimentally measured values of $S_\mathrm{x}(\omega)$ fit directly with equation~\ref{eq:sxomega}.  At high $f_\mathrm{mod}$, $S_\mathrm{x} \propto \Gamma^{+1}$ like in part (a).  For low $f_\mathrm{mod}$, however, $S_\mathrm{x} \propto \Gamma^{-1}$.  If this function is integrated with high bandwidth, the $\Gamma^{+1}$ behaviour dominates.  If integrated only out to the corner, however, $\Gamma^{-1}$ behaviour should dominate.  Expressed another way, the high $Q$ assumption overestimates the integration for small $\Delta f$, needlessly adding the area between the flat pass and the $f^{-2}$ dashed lines.

The difference becomes even more intriguing when increased driven amplitude comes into play via full dynamic range.  Figure~\ref{fig:4}c and~\ref{fig:4}d show $S_{\phi}^\mathrm{x}(\omega)$ measured noise.  In Figure~\ref{fig:4}c for 15 and \SI{40}{\torr} pressures, $a_\mathrm{driven}$ hapens to be the same value.  This makes $S_\mathrm{x}$ and $S_{\phi}^\mathrm{x}$ maintain the same relationship and the noise dependence on damping is the same as in Figure~\ref{fig:4}b.  In Figure~\ref{fig:4}d on the other hand, $a_\mathrm{driven}$ is Duffing limited causing $S_{\phi}^\mathrm{x}$ to shrink more quickly with damping than $S_\mathrm{x}$ does.  This results in $S_{\phi}^\mathrm{x}$ independent of $\Gamma$ for large bandwidths and proportional to $1/\Gamma^{2}$ at small bandwidths (\textit{cf.} equation~\ref{eq:sphixomega}.  The right hand portions of the data at different pressures and damping collapse on top of each other.  This is not a coincidence, rather it is the signature of $\SNR$ being inversely proportional to $Q$ (i.e. proportional to $\Gamma$), resulting in no $\Gamma$ dependence by Robins' equation (eqns. \ref{eq:DRformula} and \ref{eq:sigma}).  

However, consider Fig.~\ref{fig:4}d noise if integrated over bandwidth of \SI{1}{\kilo\hertz} or below.  The integration never reaches the $1/f^2$ rolloff portion of the graph.  Noise measured with this smaller bandwidth is just integrating a constant giving $\sigma^2 \propto \Gamma^{-2}$, therefore, it should result in $\sigma \propto \Gamma^{-1}$.  That is, better stability results from more damping.  Integration of white (flat) $S_\phi (\omega)$ is also known to give $\sigma \propto \tau^{-1}$ dependence \cite{Barnes1971}.  The full functional form of $\sigma$ for this case, which we refer to as the flatband regime, is (derived in SI, section 2):
\begin{equation} \label{eq:sigmatau}
\sigma_{\mathrm{fb}} (\tau) = \left ( \frac{3}{2} \right )^{1/2} \frac{1}{\SNR} \frac{1}{\Omega \tau}.
\end{equation}

This regime is not usually considered as it would normally result in prohibitively low bandwidths.  Other noise sources, such as drift, also take over close-in to carrier, often masking this regime.  However, as devices reach higher frequencies, and as $Q$ is pushed purposefully down, the corner frequency of $(\Gamma/2)/(2 \pi)$ can become very large in principle; in the present case, it is almost \SI{200}{\kilo \hertz} for atmospheric pressure.

Returning to the Allan deviation in Fig. 4, the dashed lines with $\tau^{-1}$ slope correspond to flatband theoretical minima, equation~\ref{eq:sigmatau}, for if phase noise was caused exclusively from displacement noise, and integrated over its flat region to the left of the corner frequency. The experimental data is dominated by drift or other noise sources at $\tau$ = \SI{0.1}{\second}, but generally reaches close to the theoretical limit (equation~\ref{eq:sigmatau}) of dominated by displacement noise at $\tau$ = \SI{2}{\milli\second}.

Figure~\ref{fig:5}b shows the value of Allan deviation at $\tau$ of \SI{2}{\milli\second} as a function of $Q$ along with both theoretical minimum floors from equations~\ref{eq:sigma} and~\ref{eq:sigmatau}.  It is clear that the experimental data is tracking closely to equation~\ref{eq:sigmatau} while falling well below equation~\ref{eq:sigma}.  Within the Duffing limited regime, where $\SNR \propto 1/Q$, we see that equation~\ref{eq:sigmatau} implies $\sigma \propto Q$.  Indeed, the experimental data seems to be proportional to $Q$ in this region.  Incredibly, stability gets better in proportion to the amount of damping.

\section*{Application of damping improved stability: temperature sensing}

We demonstrate an application of enhanced sensitivity with increased damping by showing temperature resolution of a NEMS beam improving with increasing pressure.  The NEMS can be used as a thermometer due to changes in resonance frequency caused by subtle temperature changes to Young's modulus and device dimensions~\cite{Inomata2016, Zhang2013}.  While traditionally in the range of \SI{-50}{ppm \per \kelvin} for silicon (ppm = parts per million), intrinsic tension changes give our devices a wide range of temperature coefficients with resonant frequency (TCRF) which can be as high as \SI{-1200}{ppm \per \kelvin} (see SI, section 1.7 and Ref.~[\!\citenum{Inomata2016}]).  The optical microring cavity itself also has a resonance dependence on temperature, primarily from the thermo-optic effect so the ring is used as a secondary temperature calibration and sensor.  The temperature responsivities for both microring and NEMS, in \SI{}{\pico\meter \per \kelvin} and \SI{}{\hertz \per \kelvin}, respectively, are simultaneously determined at each pressure tested by monitoring the change in resonant wavelength and mechanical resonant frequency for several \SI{1}{\kelvin} temperature steps (See SI, section 1.7 for details).

\begin{figure}[]
\centering
\includegraphics{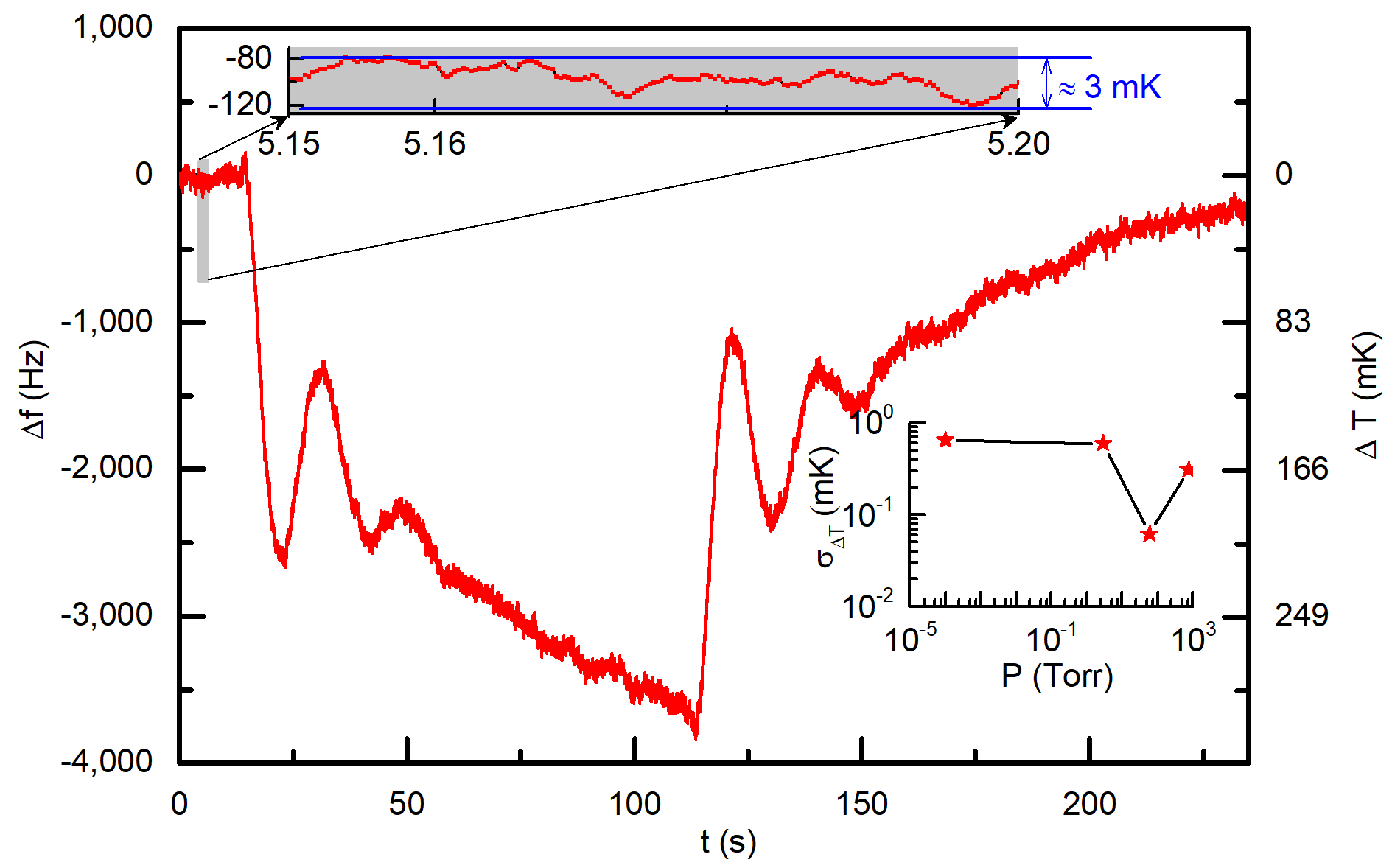}
\caption[NEMS thermometry.]{\textbf{NEMS thermometry.}  NEMS tracking of a 0.3 K underdamped-controlled temperature change turned on at about 15 s and turned off at about 115 s.  Upper inset: close-up of fluctuations in a 3 mK temperature and 50 ms time range.  Lower inset: temperature fluctuation noise-floor vs pressure. \label{fig:6}}
\end{figure}

Figure~\ref{fig:6} shows the NEMS response at \SI{3}{\torr} pressure to a \SI{+0.3}{\kelvin} step change (followed later by a \SI{-0.3}{\kelvin} step change) in the temperature controller setting.  The oscillations and long settling result from the PID controller settings combined with lag due to slight distance between the chip surface and Pt RTD temperature sensor.  The noise visible on the NEMS trace gives an idea of the minimum resolvable temperature change of the order of \SI{1}{\milli\kelvin}.  More formally, the lower inset presents the temperature resolution $\sigma_{\Delta T}$ as a function of pressure, where $\sigma_{\Delta T} = \sigma  f_0 / S_{f,T}$, and $S_{f,T}$ is the NEMS temperature responsivity of \SI[separate-uncertainty]{-12600 \pm 100}{\hertz / \kelvin}.  Data shown is for $\tau$ = \SI{5}{\milli\second} averaging time.  Similar to Fig. ~\ref{fig:5}, the NEMS temperature resolution improves with increasing pressure up to a sweet spot at \SI{60}{\torr} where it reaches \SI{50}{\micro\kelvin}.  This is comparable to references~\citenum{Inomata2016,  Zhang2013}.

\section*{Discussion}

That resolution could be independent of $Q$ in the Robins picture has been hinted at~\cite{Sansa2016, Gavartin2013}, but not tested, and not widely appreciated in the NEMS community. The further revelation that low-bandwidth sensitivity actually improves with damping is a momentous development with implications in NEMS, AFM, and other fields. As an example, the AFM community has long known of force noise proportional to square root damping, and has tried to reduce the apparent thermal force noise off resonance by increasing $Q$. This works for high bandwidth (above the corner), but increases noise on resonance, which is usually truncated and ignored. However, by purposefully suppressing $Q$, one simultaneously suppresses close-in noise while extending the corner frequency (and bandwidth). In essence, the usually inevitable tradeoff between bandwidth and low-noise is eliminated.

It is known that Eqn. 6 has no explicit $\Omega$ dependence. Equation 10, on the other hand, varies inversely with $\Omega$, opening additional paths to sensitivity improvement. Increasing the mechanical frequency should directly improve flatband sensitivity, while also extending the bandwidth available for a given $Q$. These enhancements are in addition to simultaneous sensitivity improvements coming from mass reduction.

The flatband suppression of the thermal noise peak is reminiscent of cold damping and feedback cooling~\cite{Chan2011, Teufel2011}, but is distinct in that thermal noise is spread out rather than reduced.  As such, feedback cooling could give cumulative benefit with the flatband technique. Similarly, techniques for using the nonlinear regime~\cite{Villanueva2013} or parametric squeezing~\cite{Poot2015} can be piggy-backed with flatband.

Another side-benefit of low $Q$ is suppression of intrinsic resonator frequency fluctuation noise~\cite{Sansa2016, Gavartin2013, Sun2016, Fong2012, Moser2014, Maillet2016}. Reference~\citenum{Sansa2016} recently noted this noise as ubiquitous in preventing NEMS from reaching thermal limits (though Gavartin, \textit{et al.},~\cite{Gavartin2013} were able to mitigate it with sophisticated force feedback). The transfer function responsible for conveying this intrinsic noise is proportional to $Q$~\cite{Fong2012} which may help explain why we do not see it atmospheric pressure, and see clear evidence of it only at long gate times in vacuum.

We note the limitations of our drive power keep us from accessing the full dynamic range at atmospheric pressure.  This problem can be solved by using optomechanical drive force which can be turned up almost with impunity.  Nonlinearities in the optomechanical transduction, in both readout and excitation, could eventually limit the present technique from extending dynamic range indefinitely.

\section*{Acknowledgements}

The authors acknowledge the National Research Council's Nanotechnology Research Centre and its fabrication, microscopy, and measurement facilities, Alberta Innovates Technology Futures, Alberta Innovates Health Solutions, the Natural Sciences and Engineering Research Council, Canada, and the Vanier Canada Graduate Scholarship program. The fabrication of the devices was facilitated through CMC Microsystems (silicon photonics services and CAD tools), and post processing was performed at the University of Alberta nanoFAB. We thank Paul Barclay and Mark Freeman for thoughtfully reviewing the manuscript.

\section*{Methods}
Our nano-optomechanical system is shown in Figure 1d with the principle of detection in Figure 1e. Light couples from a silicon strip waveguide to circulate in a race-track optical cavity resonator. In-plane displacement of the doubly clamped beam mechanical resonator (NEMS) modifies the local index of refraction of the racetrack, which changes the optical resonance wavelength. With the probe light parked on the side of the cavity, mechanical vibration is transduced to modulation of the optical transmission. Multiple passes of the light contributes to the excellent displacement sensitivity. Detailed analysis of the optomechanical system can be found in the SI, Section 1.4. The strength of our optomechanical coupling has been chosen strategically to resolve thermomechanical noise while still providing linear transduction to the upper end of dynamic range. The optomechanical chip is placed on a shear piezo for mechanical actuation, on a copper plate for temperature control, and is housed in a sealed chamber to allow varying the pressure (Supplementary Fig. S1). Tunable 1550nm laser light is free-space coupled through a window into and out of grating couplers on-chip. The system is controlled by a lock-in amplifier with a power amplifier providing high RF gain to drive the piezo (see SI, Section 1.1).


\putbib[DR_bibliography]
\end{bibunit}


\begin{supplementary}

\begin{bibunit}[naturemag]

\pagestyle{fancysupplementary}
\begin{addmargin}[1cm]{0cm}
\subsection*{\\  Supplementary Information: Improving mechanical sensor performance through larger damping}

\begin{addmargin}[1cm]{0cm}
Swapan K. Roy,$^{1, 2}$ Vincent T. K. Sauer,$^{1, 3}$ Jocelyn N. Westwood-Bachman,$^{1, 2}$
Anandram Venkatasubramanian,$^{1, 3}$ and Wayne K. Hiebert$^{1, 2, a)}$\\
\em{$^{1)}$Nanotechnology Research Centre, National Research Council, Edmonton, Canada\\
$^{2)}$Department of Physics, University of Alberta\\
$^{3)}$Department of Biological Sciences, University of Alberta}
\end{addmargin}
\vfill
\rule{1.7cm}{1pt}\\
$^{a)}$Corresponding author: wayne.hiebert@nrc-cnrc.gc.ca
\end{addmargin}

\pagebreak
\newpage
\pagebreak
\newpage

\section{Experimental details}

\subsection{Experimental setup}
 
\begin{figure}[h]
\centering
\includegraphics{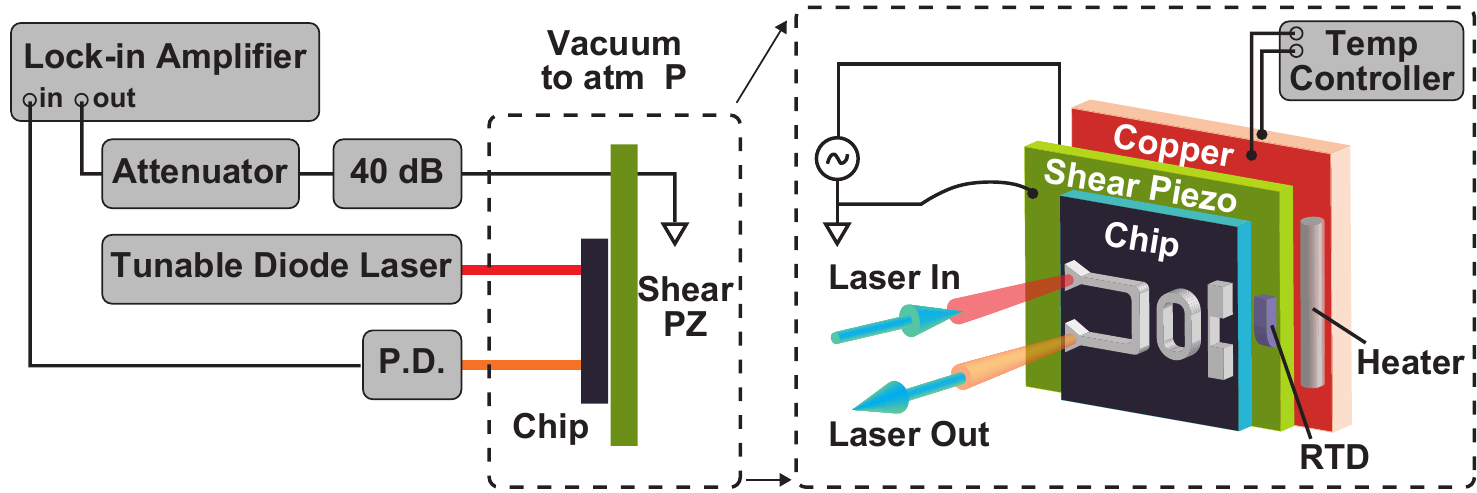}
\caption[Setup and thermal loading]{\textbf{Schematic of experimental setup.}  The right panel depicts a graphical arrangement of a NOMS device on a Silicon chip surface mounted on the top of the piezo shaker (green slab) followed by further mounting on a copper plate which facilitates a thermal contact to the base of the  pressure controlling chamber. The heater on top of copper is controlled by the temperature controller via the temperature sensor placed on the top surface of the piezo and just a few mm away from the chip edge. The left panel is the arrangement of detection and measurement unit where a Zurich instrument HF2 lock-in is the heart of nanomechanical vibration characterizations. The 40 dB box represents a Minicircuits LZY-$22+$  power amplifier, through which a maximum power of \SI{+45}{dBm} is available. In experiments, drive power is generally limited to  \SI{+36}{dBm} before the shear piezo starts to heat substantially and shift the optomechanical resonance.}
\label{s:fig:S1}
\end{figure}
  The doubly clamped beam mechanical resonance is detected using an all-pass implementation of a racetrack resonator optical cavity \cite{Diao2013,Sauer2014}. A Santec TSL-510 fiber coupled tunable diode laser (TDL) is used to probe the device. To achieve the largest displacement sensitivity, the measurement wavelength is detuned from the optical cavity center by approximately half the cavity linewidth. For both thermomechanical and driven signals, the power modulation of the detuned probe which is caused by the mechanical beam motion is measured using a Zurich Instruments HF2LI lock-in amplifier (LIA). The LIA provides the drive voltage sent to the shear-mode piezo (Noliac CSAP03) which is used to mechanically drive the DCBs in the wafer plane. A power amplifier (Minicircuits LZY-22+) is used to achieve higher drive when required. The NOMS chip is mounted on the piezo shaker with thermal conductive silver epoxy.  The piezo is placed on top of a copper plate with an attached resistive heater and platinum resistance thermometer (RTD) (both placed roughly as drawn) which are operated using a PID controller (Cryo-con Model 24C). The device is placed in a vacuum chamber, and light from the TDL is coupled from free space through the chamber's optical window and into the nanophotonic circuits using TE-mode optical grating couplers. The chamber is pumped to below \SI{1e-4}{\torr}, and a bleed valve is used to raise the pressure in the chamber to change the damping in the system.Like the Allan deviation measurements, the DCB is implemented into a phase-locked loop (PLL) using the Zurich's built-in PLL module to track any shift in resonance frequency due to temperature change made by the resistive heater.
  
\subsection{Thermomechanical noise calibration}

Accurately determining the displacement noise floor (\textit{cf}. Fig. 1c, Fig. 2, and Eqn. 3) is crucial for the analysis in this work.  We follow the standard established method for thermomechanical noise calibration \cite{Bunch2007, Li2008} which is nicely detailed in \cite{Hauer2013}.  A summary of the procedure appears below.

The voltage noise power spectral density ($S_\mathrm{{V}}$ in V$^{2}$Hz$^{-1}$) of the photodetector output, if peak shaped (as in Fig. 1c), can be assumed to be the sum of thermomechanical noise and a white background (due to instrumentation noise)
\begin{equation}
\label{s:eq:VLorentzianBG}
S_{\textup{V}}(\Omega )= S_{\textup{V}}^{\textup{th}}(\Omega )+S_{\textup{V}}^{\textup{white}}(\Omega )
\end{equation}
By comparing the measured noise to theoretically expected displacement noise spectral density $S_\mathrm{{x}}$ in m$^{2}$Hz$^{-1}$, we can calibrate the system responsivity $\Re$ in Vm$^{-1}$.  We measure $S_\mathrm{{V}}$ using a Zurich instrument HF2 lock-in amplifier in zoomFFT mode up to 78 Torr and by an Agilent 8593E spectrum analyzer from $120-760$ Torr (the latter being better suited to larger frequency spans) while holding ambient temperature constant at 298 K. Measured peaks and quality factors ($Q$) are used in the calibration.  What is needed is a theoretical functional form for $S_\mathrm{{x}}=S_\mathrm{{x}}^{\textup{th}}+S_\mathrm{{x}}^{\textup{white}}$.  This is derived via equipartition theorem (\textit{cf}. section 2.1) resulting from the Langevin (random thermal) force acting on the resonating normal mode and is given by
\begin{equation}
\label{s:eq:S1}
S_{\textup{x}}^{\textup{th}}(\Omega )=\frac{S_{\textup{F}}^{\textup{th}}(\Omega )}{M_{\textup{eff}}^{2}}\frac{1}{(\Omega _{\textup{0}}^{2}-\Omega ^{2})^{2}+(\Gamma \Omega )^{2}}= \frac{S_{\textup{F}}^{\textup{th}}(\Omega )}{M_{\textup{eff}}^{2}}\frac{1}{((\Omega _{\textup{0}}-\Omega)(\Omega _{\textup{0}}+\Omega))^{2}+ (\frac{\Omega _{\textup{0}}}{Q}\Omega )^{2}}
\end{equation}
where, $S_{\textup{F}}^{\textup{th}}= \frac{4k_{\textup{B}}TM_{\textup{eff}}\Omega _{\textup{0}}}{Q}$ in $\textup{N}^{2}\textup{Hz}^{-1}$ is the thermal force spectral density acting on the nanoscale resonator. Here, $k_{\textup{B}}$, $M_{\textup{eff}}$, $\Omega _{\textup{0}}/2\pi$, $Q$  and,  $\Gamma/2\pi =\frac{\Omega _{\textup{0}}}{Q}/2\pi $  are Boltzmann constant, effective mass, resonance frequency, quality factor and linewidth of the DCB resonator. At  $\Omega =\Omega _{\textup{0}}$ equation \ref{s:eq:S1} reduces to 
\begin{equation}
\label{s:eq:S2}
S_{\textup{x}}^{\textup{th}}(\Omega_{\textup{0}} )= \frac{4k_{\textup{B}}TQ}{M_{\textup{eff}}\Omega _{\textup{0}}^{3}} 
\; \textup{m}^{2}\textup{Hz}^{-1}
\end{equation}
Thus the r.m.s displacement peak of the power spectral density in absence of any background noise can be found as (in a 1 Hz bandwidth)
 \begin{equation}
 \label{s:eq:S3}
a_{\textup{th}}=\sqrt{S_{\textup{x}}^{\textup{th}}(\Omega_{\textup{0}})}  \; \textup{m}\; \textup{Hz}^{-\frac{1}{2}}\times \textup{1 Hz}^{\frac{1}{2}} = \sqrt{\frac{4k_{\textup{B}}TQ}{M_{\textup{eff}}\Omega _{\textup{0}}^{3}} } \; \textup{m}
\end{equation}
If $|\Omega_0-\Omega|\ll \Omega_0$,  then the displacement spectral density curve described in equation \ref{s:eq:S1} can be reduced with approximations $ (\Omega _{\textup{0}}-\Omega)(\Omega _{\textup{0}}+\Omega) \cong 2\Omega _{\textup{0}}(\Omega _{\textup{0}}-\Omega)$ \; and\;   $\frac{\Omega _{\textup{0}}}{Q}\Omega \cong \frac{\Omega _{\textup{0}}}{Q}\Omega _{\textup{0}}$ as below
\begin{equation}
\label{s:eq:LorentzianSX}
S_{\textup{x}}^{\textup{th}}(\Omega )= \frac{S_{\textup{F}}^{\textup{th}}(\Omega )}{M_{\textup{eff}}^{2}}\frac{1}{4\Omega _{\textup{0}}^{2}(\Omega _{\textup{0}}-\Omega)^{2}+ (\frac{\Omega _{\textup{0}}}{Q}\Omega _{\textup{0}} )^{2}}= \frac{1}{\Omega _{\textup{0}}^{2}}\frac{S_{\textup{F}}^{\textup{th}}(\Omega )}{M_{\textup{eff}}^{2}}\frac{1}{4(\Omega _{\textup{0}}-\Omega)^{2}+ (\frac{\Omega _{\textup{0}}}{Q})^{2}}
\end{equation}
Equation \ref{s:eq:LorentzianSX} is a Lorentzian function to which a white background can be added
\begin{equation}
\label{s:eq:LorentzianBG}
S_{\textup{x}}(\Omega )= S_{\textup{x}}^{\textup{th}}(\Omega )+S_{\textup{x}}^{\textup{white}}(\Omega )
\end{equation}
By fitting the voltage noise to a Lorentzian with background (directly comparing equation \ref{s:eq:VLorentzianBG} with equations \ref{s:eq:LorentzianSX} and \ref{s:eq:LorentzianBG}), the calibration of $S_{\textup{x}}$ to $S_{\textup{V}}$ is naturally achieved.
\subsubsection{Calculation of displacement responsivity, $\Re\; \textup{Vm}^{-1}$ }
A Lorentzian curve fit was performed for measured  $ S_{\textup{V}} \; \textup{V}^{2}\textup{Hz}^{-1}$  at each pressure to obtain the resonance frequency, $f_{0}$ and mechanical quality factor, $Q$ and the background $ S_{\textup{V}}^{\textup{white}}$.  The peak height of this measured spectral density can be calculated as
\begin{equation}
\label{s:eq:SVth}
   S_{\textup{V}_{\textup{pk}}}^{\textup{th}} = S_{\textup{V}}(\Omega_{\textup{0}})-S_{\textup{V}}^{\textup{white}} \; \textup{in }\; \textup{V}^{2}\textup{Hz}^{-1}	
\end{equation}
Now, plugging the measured $f_{0}$  and  $Q$ from Lorentzian fit into equation \ref{s:eq:S2} gives displacement power spectral density $S_{\textup{x}}^{\textup{th}}(\Omega_{\textup{0}} )\; \textup{in} \; \textup{m}^{2}\textup{Hz}^{-1}$ of the resonator vibration at its resonance frequency and depends on damping induced by the chosen pressure.  Defining $S_{\textup{x}_{\textup{pk}}}^{\textup{th}}$  as 

\begin{equation}
\label{s:eq:SXth}
   S_{\textup{x}_{\textup{pk}}}^{\textup{th}} = S_{\textup{x}}(\Omega_{\textup{0}})-S_{\textup{x}}^{\textup{white}} \; \textup{in }\; \textup{m}^{2}\textup{Hz}^{-1}	
\end{equation} 
means that $\sqrt{S_{\textup{x}_{\textup{pk}}}^{\textup{th}}} \; \textup{in}\;\textup{ m}\; \textup{Hz}^{-\frac{1}{2}}$ must be equal to the measured peak height,$ \sqrt{S_{\textup{V}_{\textup{pk}}}^{\textup{th}}} \; \textup{in}\;\textup{ V}\; \textup{Hz}^{-\frac{1}{2}}$ of voltage spectral density given by equation \ref{s:eq:SVth}. Thus, measued voltage in experiments can easily be converted into displacement by obtaining the conversion factor, $ \Re $ as below
\begin{equation}
\label{s:eq:Responsivity}
 \Re\; \textup{Vm}^{-1}=\frac{\sqrt{S_{\textup{V}_{\textup{pk}}}^{\textup{th}}} \;\textup{ V}\; \textup{Hz}^{-\frac{1}{2}}}{\sqrt{S_{\textup{x}_{\textup{pk}}}^{\textup{th}}} \;\textup{ m}\; \textup{Hz}^{-\frac{1}{2}}}  
\end{equation}
Figure 1c and Fig. 2 use this method to calibrate the vertical axis.  

\subsubsection{Background noise floor}
The possible sources of background noise in our nanophotonic detection system are the Johnson noise of electronic measurement instruments e.g. HF2 lock-in or spectrum analyzer ($ 5 \; \textup{nV}\textup{Hz}^{-\frac{1}{2}}$from instrument manual), shot noise, $ S_{\textup{V}}^{\textup{shot}}$ from laser source and dark current $ S_{\textup{V}}^{\textup{dark}}$  of the photodetector.  The total background is the sum  of these $S_{\textup{V}}^{\textup{white}}=S_{\textup{V}}^{\textup{elec}}+S_{\textup{V}}^{\textup{shot}}+S_{\textup{V}}^{\textup{dark}}$. Measured optical power to voltage conversion factor for a $ 50\; \Omega$ termination is \cite{Diao2013}, $\mathfrak{O}= 15\; \textup{V}\textup{mW}^{-1} = 15000\; \textup{V}\textup{W}^{-1} $. The free space optical beam shot noise is defined as 
 \begin{equation}
S_{\textup{opt}}^{\textup{shot}}= 2h\nu \left \langle P \right \rangle
 \end{equation}
where the Planck's constant $ h= 6.64\times 10^{-34}\;  \textup{m}^{2}\textup{kgs}^{-1}$ ;  the laser frequency, $ \nu =c\lambda ^{-1}= 1.93\times 10^{14}\; \textup{Hz} $ for $1550 \; \textup{nm}$  wavelength; from the DC transmission data the average power, $\left \langle P \right \rangle= \frac{T_{\lambda \textup{probe}}}{\mathfrak{O}}\; \frac{\textup{V}}{\textup{VmW}^{-1}}\approx \frac{0.08}{15}=0.0053\; \textup{mW} $.
 
With the detector quantum efficiency, $\eta $ the power spectral density at the photodetector can be found as follows
\begin{equation}
\label{s:eq:shot}
S_{\textup{W}}^{\textup{shot}}=\frac{2h\nu \left \langle P \right \rangle}{\eta }\; \textup{W}^{2}\textup{Hz}^{-1}
\end{equation}
where, $\eta =\frac{R_{\lambda }}{\lambda }\times \frac{hc}{e}=\frac{1\; \textup{AW}^{-1}}{1550\;\textup{ nm}}\times 1240\frac{\textup{Wnm}}{\textup{A}}=0.8$. Now plugging all values in equation \ref{s:eq:shot} we have, $\sqrt{S_{\textup{W}}^{\textup{shot}}}= 1.3 $  $\textup{pWHz}^{-1}$ which gives the power spectral density of shot noise in voltage by $\sqrt{S_{\textup{V}}^{\textup{shot}}}=\sqrt{{S_{\textup{W}}^{\textup{shot}}}}\times \mathfrak{O}= 19.5\; \textup{nV}\textup{Hz}^{-\frac{1}{2}}$

After blocking all input light, the measured dark current, $\sqrt{S_{\textup{V}}^{\textup{dark}}}$ of photodetector around the resonance frequency from Zurich lock in amplifier is found as $196\; \textup{nVHz}^{-\frac{1}{2}}$ and from spectrum analyzer as $126\; \textup{nVHz}^{-\frac{1}{2}}$. This results in $(S_{\textup{V}}^{\textup{white}})^{1/2}$ of $197\; \textup{nVHz}^{-\frac{1}{2}}$ and $128\; \textup{nVHz}^{-\frac{1}{2}}$ for vacuum and atmospheric pressure, respectively.  Expressed in displacement noise (converted using responsivity (equation \ref{s:eq:Responsivity})) $(S_{\textup{x}}^{\textup{white}})^{1/2}$ is 
$\approx 20.3$  for lock-in and $\approx 13.1$ for spectrum analyzer  in $\textup{fmHz}^{-\frac{1}{2}}$.

As described in Ref.~\citenum{durig1997dynamic}, this incoherent background noise floor sets an ultimate limit to the frequency stability as follows
\begin{equation}
\frac{\delta f_{\textup{background}}}{f_{\textup{0}}}=\frac{(S_{\textup{x}}^{\textup{white}})^{1/2}}{\pi f_{\textup{0}}x_{\textup{driven}}}\tau ^{-3/2}
 \label{s:eq:ADbackground}.
\end{equation}
where $x_{\textup{driven}}$ is driven amplitude and $\tau$ is a sampling time.  Equation~\ref{s:eq:ADbackground} is plotted in Fig. 4 as the orange shaded region in the lower left corners (it is within the plotted range only for vacuum and atmospheric pressures).


\subsection{Determination of onset of nonlinearity}

Accurately determining the onset of nonlinearity is important for defining the upper cutoff of the dynamic range. In this section, we describe the calibration of onset of nonlinearity in the doubly clamped beams.

Spatial shift of NEMS resonance frequency with increasing vibration amplitude is a well-known phenomenon \cite{Duffing1918, Kovacic2011, Andres1987, Nayfeh2008, Postma2005, Kacem2009, Schmid2016}. When external driving power is increased enough, vibration amplitude no longer increases linearly. Similar to rf-electronics, the resonance mode of the NEMS enters into a non-linear regime where hysteresis and gain compression occur. The maximum amplitude where linear response ends is often referred as the onset of nonlinearity or critical amplitude, $a_{\textup{c}}$. Above critical amplitude, the vibrating mechanical element experiences various nonlinearities in its restoring force, e.g., elongation of the beam, defects in clamping, material nonlinearity, existence of any force gradient in the system due to detection or actuation or even thermal gradient. In our DCB resonators, strain induced tension, geometrical nonlinearity occurs. This can be described by the Duffing equation by introducing a cubic nonlinearity term in the second-order differential equation of simple harmonic motion\cite{Postma2005, Kacem2009, Schmid2016}.

The critical amplitude $a_{\textup{c}}$ occurs when the frequency solution to the Duffing equation just starts to be multivalued (\textit{i.e.} the bifurcation point) and is characterized by a section of infinite slope and the start of hysteresis in frequency sweeps. In Postma \textit{et al.}\cite{Postma2005} the expression for critical amplitude, $a_{\textup{c}}$ is given as (when considering no residual tension in the DCB resonator)
\begin{equation} \label{s:eq:ac1}
a_{\textup{c}}= \Omega _{\textup{0}}\frac{L^{2}}{\pi ^{2}}\sqrt{\frac{\rho \sqrt{3}}{EQ}}
\end{equation}
where, $\Omega _{\textup{0}}$ is the resonance frequency of the DCB resonator with a length $L$. $\rho$ and $E$ are the density and Young's modulus of the material. Here, $Q$ is the measured quality factor of the resonator. In a doubly clamped beam with a residual tension \cite{Postma2005},$T_{\textup{0}}$, the onset of nonlinearity is as below
 \begin{equation} \label{s:eq:ac2}
a_{\textup{c}}= \frac{2}{\sqrt[4]{3}}\sqrt{\frac{1}{Q}\left ( \frac{d^{2}}{3}+\frac{T_{\textup{0}}L^{2}}{\pi ^{2}Etd} \right )}
\end{equation}
Here, $t$ is the thickness and $d$ is the width of the beam in the direction of motion. The second term within the bracket corresponds to resonance frequency. From equation \ref{s:eq:ac2} one can tell that $a_{\textup{c}}$ increases with increasing damping (decreasing $Q$) for a particular device geometry.

In the main manuscript, the critical amplitude equation without tension is used for simplicity.  As can be seen in Figure \ref{s:fig:acrit1}, the difference between the two equations is very small.  Strictly speaking, we define $a_{\textup{crit}} =0.745 a_{\textup{c}}$ to correspond with the theoretical amplitude for 1 dB of compression, and define it as the practical end of the linear range \cite{Postma2005}.

Determination of the 1dB compression of critical amplitude, $a_{\textup{crit}}$, is done by collecting the amplitude response on resonance while sweeping driving power voltage as shown in figure \ref{s:fig:3onset} ( blue open symbol). From the linear portion of this experimental plot, a 1 dB compression line (red line) is plotted. The intersection gives the 1 dB compression of driving power or critical driving voltage, $V_{\textup{crit}}$ before the onset of nonlinearity.

 \begin{figure}[h]
\centering
\includegraphics{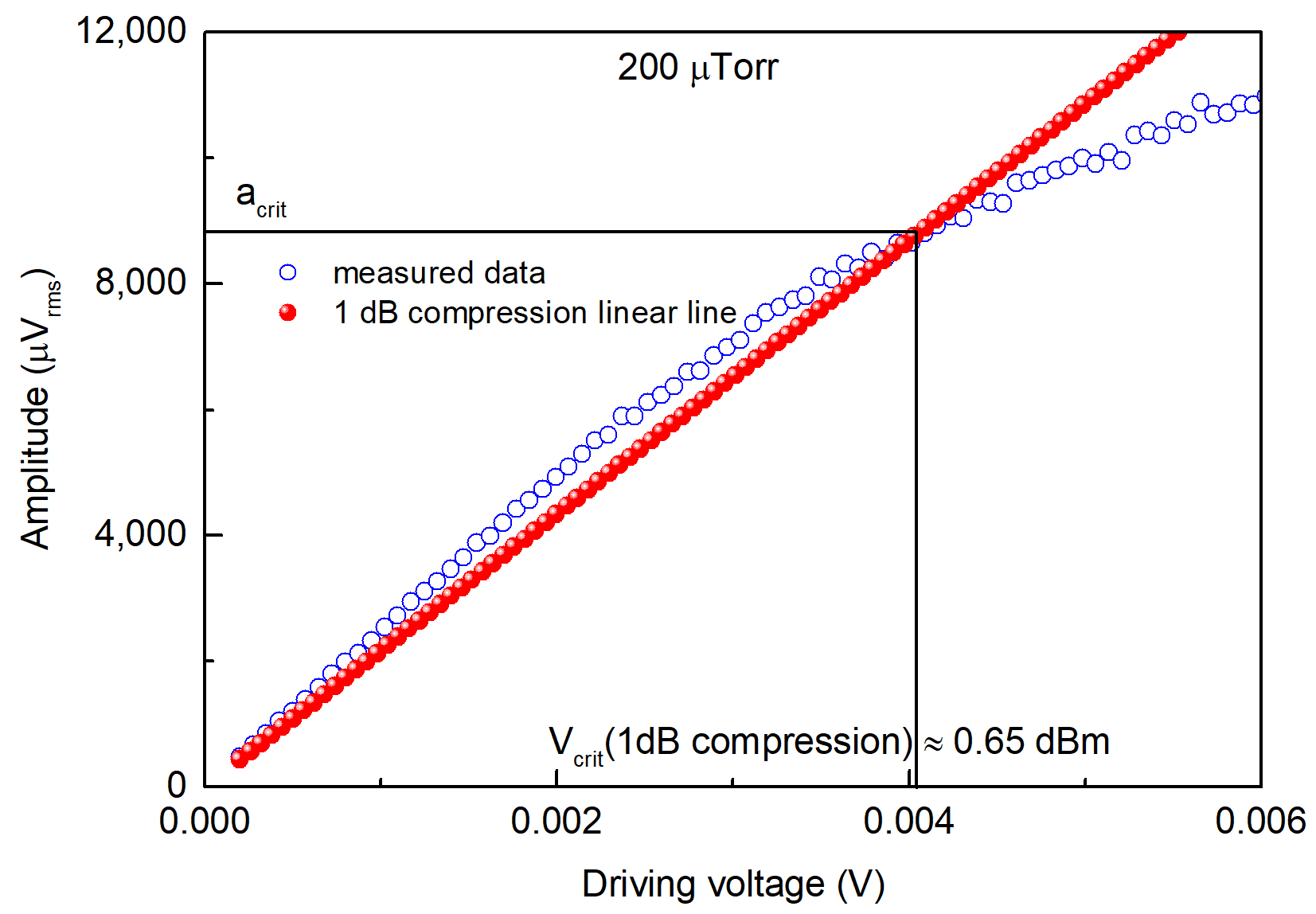}
\vspace{0.5 em}
\includegraphics{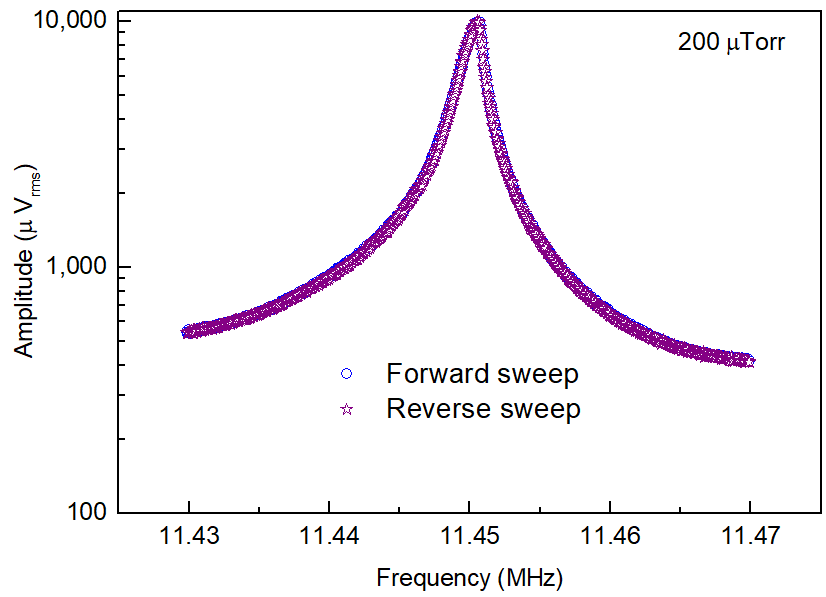}
\caption[Figure S2 short description]{\textbf{Left: A representative plot for determining critical drive power}. The onset of nonlinearity is at $200$ $ \mu \textup{Torr}$. The effective gain of the rf-amplifier is 38.3dB. The device can be driven up to 30 dBm without noticeable heating. The $\textup{1}\; \textup{dB}$ compression of critical drive power obtained for this pressure is 0.65 dBm provided by a back calculation from the crossover point of 1 dB compression line in the horizontal axis. \textbf{Right: At this critical drive of 0.65 dBm forward and reverse sweeps are not showing any hysteresis.} The measured $Q$ from Lorentzian fit of TM noise is $8286 \pm 19$ and that from the driven response in this figure $8681 \pm 460 $ from the phase slope at resonance; which means that $Q$ is independent of driving power up to the 1 dB compression of the onset of nonlinearity. The absence of hysteresis and similar $Q$ values at the driven response compared to un-driven $Q$ indicate that the device can be operated at its maximum linear amplitude.}
\label{s:fig:3onset}
\end{figure}
A forward and reverse frequency sweep at critical drive confirms that the resonance shape is just starting to tilt and hysteresis has not yet set in.  The $Q$-factor also remains similar to that measured in the thermomechanical noise.

All experimental $a_{\textup{crit}}$, from high vacuum to atmospheric pressure are compared to corresponding theoretical values given by equation \ref{s:eq:ac1} and \ref{s:eq:ac2} and plotted together with experimental values in Fig. \ref{s:fig:acrit1}. From a comparison between experimental and analytical values in the figure \ref{s:fig:acrit1} it can be inferred that the DCB beam used is subjected to geometrical nonlinearity. 
 \begin{figure}
    \centering
    \includegraphics[scale=1]{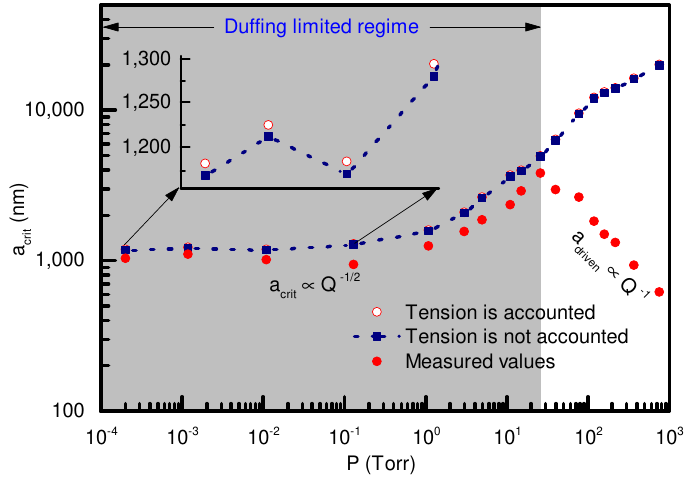}
    \caption{ \textbf{Evolution of onset of nonlinearity with increasing damping or decreasing $Q$}} 
    \label{s:fig:acrit1}
  \end{figure}

\subsubsection{Non-linearity onset: modification at high pressures}

    It is evident from equation \ref{s:eq:ac2} that for a given device (geometry is constant) with increasing damping (i.e., decreasing $Q$ ) $a_{\textup{c}}$ increases.  At the same time, decreasing $Q$ requires large chip surface motion to achieve the same amplitude, since $a_{\textup{NEMS}} \approx Q a_{\textup{surface}}$.  This combination necessitates quickly ramping up the drive power at high damping.  Higher driving power by piezo-actuation generally causes on-chip heat generation as more power is dumped into the piezoelectric. Induced heating from actuation and detection is a familiar phenomenon in NEMS. It can happen either by the heating effect of driving or by optical adsorption and is common to optomechanical devices \cite{Song2014,Meenehan2014}.  Temperature induced changes to both the resonance frequency and the ring responsivity can complicate the nonlinearity measurement when there is significant heating during the ramp in power.
    \begin{figure}[h]
    \centering
    \includegraphics[scale=1]{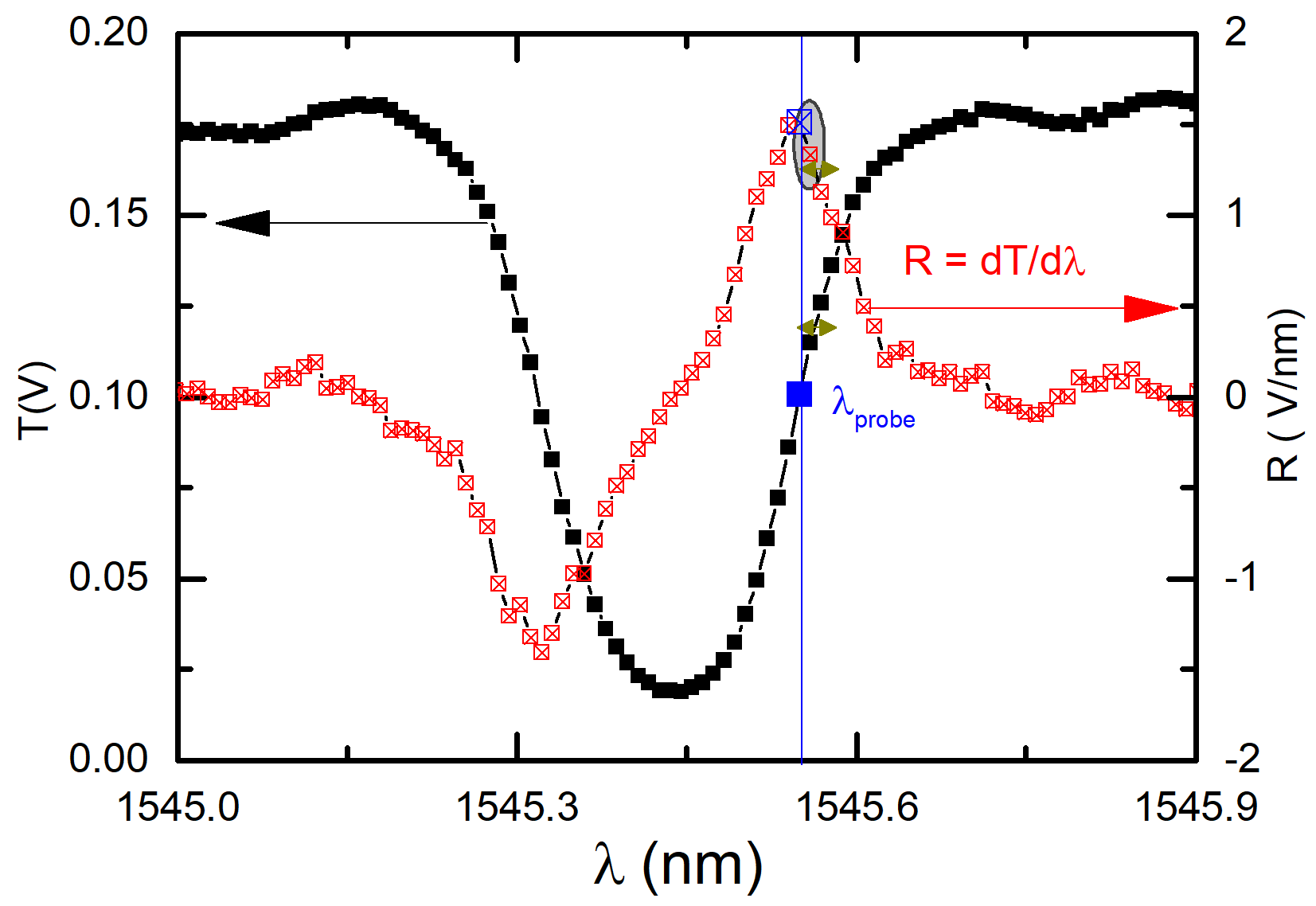}
    \caption{ \textbf{Optical resonance at $26\; \textup{Torr}$}. Left axis is the measured transmission in $ \textup{Volts} $ and the right axis is the corresponding slope. Blue data point at $1545.549 \; \textup{nm} $  has the maximum slope, and probe wavelength is set at this wavelength for a transmission power around $0.1\; \textup{Volts}$. By sitting on probe wavelength we are able to collect any transient change in probe power (transmission) by a home-built lab-view program. Dark yellow arrow symbol at $1545.569 \; \textup{nm}$, $0.12 \; \textup{Volts}$ is the observed experimental shift due to piezo-heating effect during the $26 Torr$ power sweep shown in the next figure. From material properties it is discussed that optical ring resonance shifts by $80\; \textup{pm}$ for $1\; \textup{K}$ temperature change . Hence, this $20\; \textup{pm}$ shift corresponds to about $0.25 \; \textup{K}$ temperature rise. The red squares are the change in slope of the optical resonance. The small gray circle shows the change in slope within the piezo-heating regime.}
   \label{s:fig:S6}
  \end{figure}

The changing responsivity is the dominant effect of the two.  Figure \ref{s:fig:S6} shows the photodetector transmission in vicinity of the ring resonance and the slope $\frac{dT}{d\lambda}$ which is proportional to the transduction responsivity.  During temperature changes, the curves shift causing transmission and responsivity changes.  It is straightforward to track these values during a power sweep, which allows correcting 1 dB compression point values.  Figure \ref{s:fig:S7}a shows photoreceiver transmission captured during vacuum, 5, 10, and 26 Torr power sweeps.  Transmission (and implied responsivity) are constant for vacuum, 5, and 10 Torr.  These sweeps max out below +30 dBm power.  For 26 Torr the power sweep goes up to +38 dBm and is accompanied by significant heating.  The experiment is conducted a few degrees above room temperature with the chip holder temperature locked by PID control.  The placement of the Pt RTD sensor directly on the piezo produces a counter intuitive effect of actually lowering the chip surface temperature as the piezo dissipates more power (this is because the PID) heater shuts off to compensate).  Thus the piezo heating blue shifts the optical ring resonance causing an increase in transmission, and a corresponding decrease in responsivity.
 \begin{figure*}[ht]
    \centering
    \includegraphics[scale=0.9]{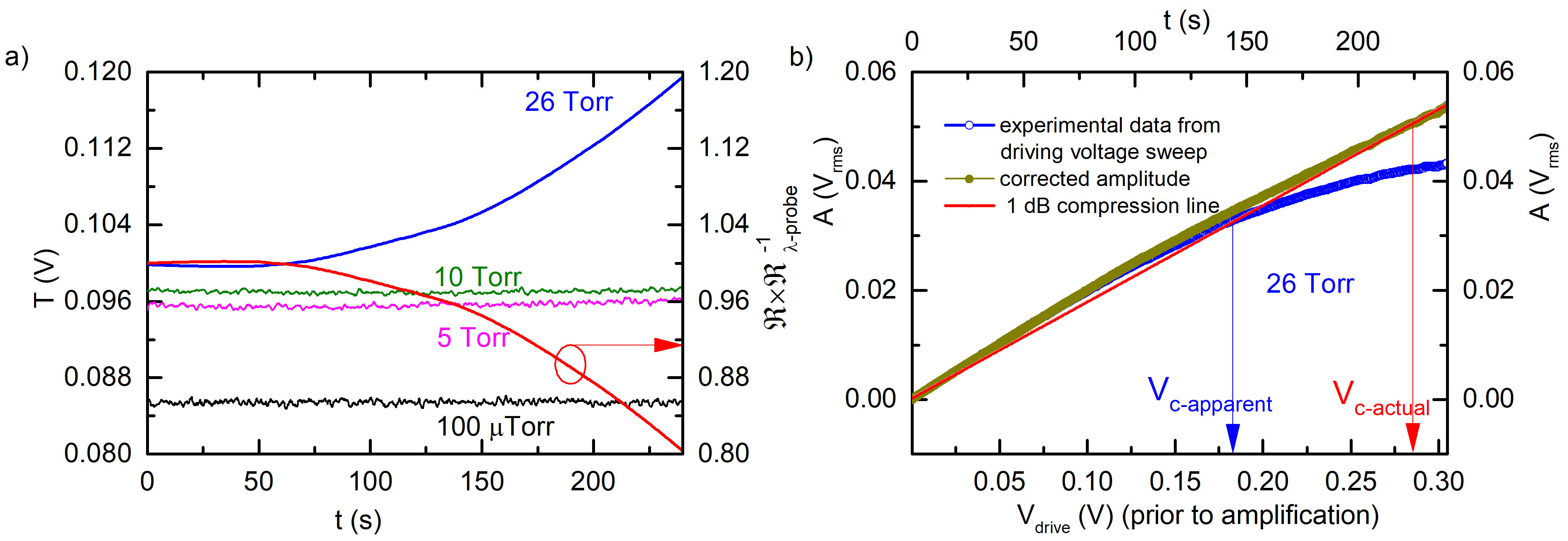}
    \caption{\textbf{Left: Left axis represents the temporal change of probe power at different pressures during voltage sweep shown in the right plot.} At low pressures flat optical transmission plots indicate absence of appreciable piezo-heating. $26$ Torr data (blue) shows a significant change due to piezo-heating with a $0.25$ $C$ temperature change. Corresponding slope change is normalized along the right axis. The slope change can be used to re-normalized data in the right panel.  \textbf{Right : Amplitude sweep and corrected amplitude sweep with increasing driving voltage.} The corrected amplitude is obtained by dividing the experimental data by the red plot in the left panel ($\frac{R}{R_{\lambda -probe}}$).
    The blue arrow indicates the early nonlinearity ( for $1$ \textit{dB} compression) at $0.188$ \textit{Volts} in lock-in tab which corresponds to $0.132$ \textit{rms Volts} or $-4.7$ \textit{dBm}.The effective gain of the rf-amplifier with a $6$ \textit{dB} attenuator is $38.3$ \textit{dBm}. Thus the apparent critical drive power from experimental data is $33.6$ \textit{dBm}. From the corrected amplitude response the actual critical drive is around $0.285$ \textit{Volts} or $0.202$ \textit{rms Volts} or $37.42$ \textit{dBm} as shown by red arrow.}
    \label{s:fig:S7}
  \end{figure*}

Figure \ref{s:fig:S7}b shows the 26 Torr power sweep plotted as response \textit{vs.} $V_{\textup{drive}}$.  The original response voltage, and the corrected response voltage (the latter divided by normalized responsivity $ \Re/\Re_{\lambda \textup{-probe}} $) give apparent and corrected critical drive values, respectively.

\subsection{Notes on optomechanics}

\subsubsection{Optomechanical coupling coefficient calibration}

The device under test is a doubly clamped beam (DCB) approximately \SI{9.75}{\micro \meter} long and \SI{160}{\nano \meter} thick in the direction of oscillation. It is fabricated on a standard nanophotonic silicon on insulator wafer with a \SI{220}{\nano\meter} thick device layer. The DCB oscillates in the plane of the wafer towards and away from a racetrack resonator optical cavity, in an all-pass configuration, which is fabricated \SI{120}{\nano\meter} away. The waveguide which creates the racetrack resonator is \SI{430}{\nano\meter} wide. The racetrack resonator has an optical $Q$ of $\sim8400$, a linewidth of \SI{0.18}{\nano\meter}, a free spectral range of $\sim\SI{13.1}{\nano\meter}$, and a finesse of $\sim70$. 

To calculate the optomechanical coupling coefficient ($g_\mathrm{om} = \partial \omega / \partial x$) from simulation, we can use the change in effective index over distance to calculate the optomechanical coupling~\cite{Li2008, Sauer2014}. This calculation results in an optomechanical coupling coefficient $g_\mathrm{om} \sim \SI{2.86}{\radian\giga\hertz\per\nano\meter}$.

The measured optomechanical devices are designed to operate deep in the Doppler regime where the overall optical cavity intensity decay rate ($\kappa$) is much, much greater than the mechanical frequency of the device ($\Omega_0$) \cite{Aspelmeyer2014}. In this way, gains are made with mechanical transduction sensitivity while minimizing optomechanical effects such as optical damping or amplification. This maintains a more simple system for a more robust sensor. The $\kappa$ of our optical racetrack is approximately \SI{1.5e5}{\mega\hertz \cdot \radian} compared to $\Omega_0 = \SI{70.3}{\mega\hertz \cdot \radian}$, which satisfies the $\kappa >> \Omega_0$ criterion. 

To confirm that the optical damping effects are negligible compared to the mechanical damping in the system, the optical spring effect is used to extract the light enhanced optomechanical coupling strength, $g$, of the system using the equation\cite{Aspelmeyer2014}
\begin{equation}
\left. \delta \Omega_0(\Delta)\right|_{\kappa >\! \!> \Omega_0} = g^2 \frac{2\Delta}{\kappa^2/4 +\Delta^2}.
\end{equation}
Above, $\Delta$ is the wavelength detuning of the probe in relation to the optical cavity centre (red-detuned: $\Delta <0$, blue-detuned: $\Delta > 0$). The measurement is taken at the greatest slope of the DC optical transmission curve on the blue and red side of the optical cavity (inset figure~\ref{s:fig:spring}) which is approximately equal to a detuning of $\pm \kappa/2$, respectively. Assuming the optical spring effects are equal and opposite for the blue and red measurement, $\delta \Omega_0 \approx \SI{3.2}{\kilo\hertz \cdot \radian}$ as shown in figure~\ref{s:fig:spring}. This gives a value of $g \approx \SI{16}{\mega\hertz \cdot \radian}$. To convert this to the optomechanical coupling coefficient for comparison to simulated values, we can use the following equation:

\begin{equation}
    g_\mathrm{om} = \frac{g}{n_\mathrm{cav}^{1/2} x_\mathrm{ZPF}}
\end{equation}

\noindent In the above equation, $n_\mathrm{cav}$ is the number of photons in the optical cavity and $x_\mathrm{ZPF}$ is the zero point fluctuations of the DCB. This results in an experimental $g_\mathrm{om} \sim \SI{2.83}{\radian\giga\hertz\per\nano\meter}$.

\begin{figure}[ht]
\centering
\includegraphics{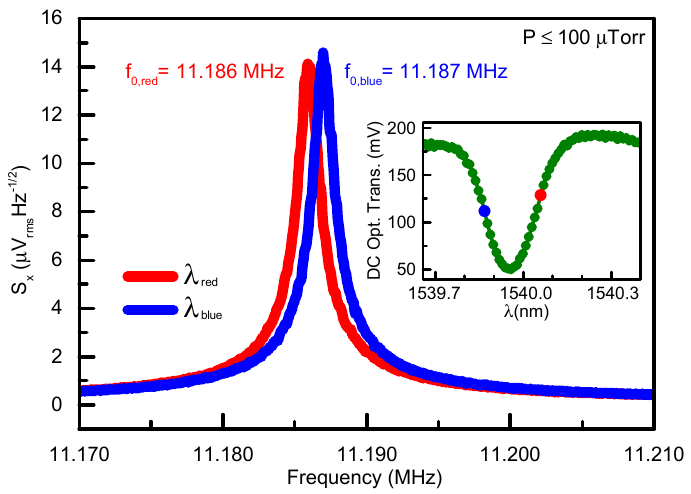}
\caption[]{\textbf{Optomechanical spring effect in the device.} Blue and red detuning amounts are shown on the inset. They are approximately at $+/-\kappa/2$, respectively, which is where maximum frequency detuning would occur. \label{s:fig:spring}}
\end{figure}

Maximum cooling/heating for the Doppler regime will occur with the detuning used in this measurement, and the maximum optical damping/amplification is calculated using\cite{Aspelmeyer2014}
\begin{equation}
\Gamma_\mathrm{opt}\left(\Delta = -\frac{\kappa}{2} \right ) = 8 \left(\frac{g}{\kappa} \right )^2 \Omega_0.
\end{equation}
This gives a value of \SI{6}{\hertz \cdot \radian} which is much less than the mechanical damping of $\sim\SI{2}{\kilo\hertz}$. This confirms that the total damping will be dominated by the mechanical element, and optomechanical damping effects can be considered negligible.

\subsubsection{Optomechanical Nonlinearity}

One potential source of nonlinearity in optomechanical systems is a readout nonlinearity. This is caused by the Lorentzian lineshape of the optical cavity. If the amplitude of the mechanical device is sufficiently large to shift the cavity out of the linear section on the side of the Lorenztian optical resonance, nonlinearities in the transduction can occur. Briefly, the nonlinearity coefficient can be calculated using the optical cavity properties and the optomechanical coupling coefficient. By starting from the expression for the dispersive optical force,
\begin{equation}
F = \frac{-2 P_\mathrm{in} \gamma_\mathrm{ex} G}{\omega (\Delta + \gamma)}
\end{equation}

\noindent and expanding about the static position $x_0$ of the mechanical resonator, we can extract the cubic spring constant $k_3$. This can be used to derive the nonlinearity coefficient $\alpha$ and therefore the critical amplitude. This calculation is explored more thoroughly in~\cite{Li2012,Bachman2018}. The minimium  critical amplitude calculated given our optical cavity parameters is 28~nm, significantly above the nonlinear amplitude observed in experiment. For this reason, we are confident that the nonlinearity is not a result of a transduction nonlinearity. 

\subsection{Acoustic interference during piezoactuation}
Large driving power and small quality factor, as we have in case of atmospheric pressure in our NOMS devices, can lead to bulk acoustic related complications in device piezoactuation.  This issue has been well summarized in the thesis of Igor Bargatin \cite{Bargatin2008} and is discussed in this section.
   In our nano-photonic measurement system we can actuate NOMS either optically or piezoelectrically \cite{Sauer2017}. With our moderate values for optomechanical constants in these devices, we have found that optical forces are insufficient to drive up to the onset of Duffing non-linearity.  Piezoshaker actuation with the aid of an rf-amplifier can provide enough driving power to test the Duffing behaviour of our devices up to $\approx 30$ Torr.
 
  We follow the usual practice in piezodrive in which the chip containing vibrating elements like NEMS (see Fig. \ref{s:fig:S1}) is glued to the top of a piezoshaker. When the piezoshaker is subjected to driving voltage it physically shakes the chip containing NEMS devices.  The amplitude of the chip surface motion, $ a_{s} $, applies a center of mass force to the NEMS of $F_{in} = M_{eff}\Omega _{0}^{2}a_{s}^{2}$, where $ M_{eff}$ and $ \Omega _{0}/2\pi$ are the effective mass and resonance frequency of the device in vibration. In the ideal scenario $ a_{s} $, is assumed frequency independent (i.e. uniform within the frequency sweep range).  For a high $Q$ device (which has a "narrow" frequency span) amplitude of this surface motion is negligible compared to the resonator's amplitude $ a_{NEMS} $. If $Q> > 1 $, the amplitude of the NEMS can be written as
  \begin{equation}  \label{s:eq:aisQa}
 a_{NEMS}=  Q \times a_{s}
 \end{equation}

For frequencies over 1 MHz, $ a_{s} $ is not uniform across the surface and varies by frequency for a given applied RF driving voltage. Propagation of ultrasonic waves inside the piezoshaker and NEMS substrate, including interface reflections, can result in complicated interference patterns of these waves. A complex spatial and frequency dependent motion of the chip surface due to such bulk acoustic interference results in frequency dependent drive strength (i.e. $ a_{s} $). This results in a forest of weak, bulk-acoustic related resonance peaks when a large frequency is spanned. Depending on the size of the piezoshaker and the chip mounted on it, there is a characteristic span of driving frequency, $\Delta f$, within each acoustic resonance where the surface motion may be considered quasi-uniform. This $\Delta f$ can vary at different frequencies. If a high $Q$ NEMS is driven within any of the $\Delta f$, the NEMS resonance can be described by equation \ref{s:eq:aisQa} because of negligible and quasi-uniform magnitude of $ a_{s} $ compared to $ a_{NEMS} $ . In larger damping, when $ \Gamma > > \Delta f$, then resonance shape of the NEMS can be severely distorted (\textit{cf} Fig. 2 for $40$ and $760$ Torr).
 
 \begin{figure}[ht]
\centering
\includegraphics{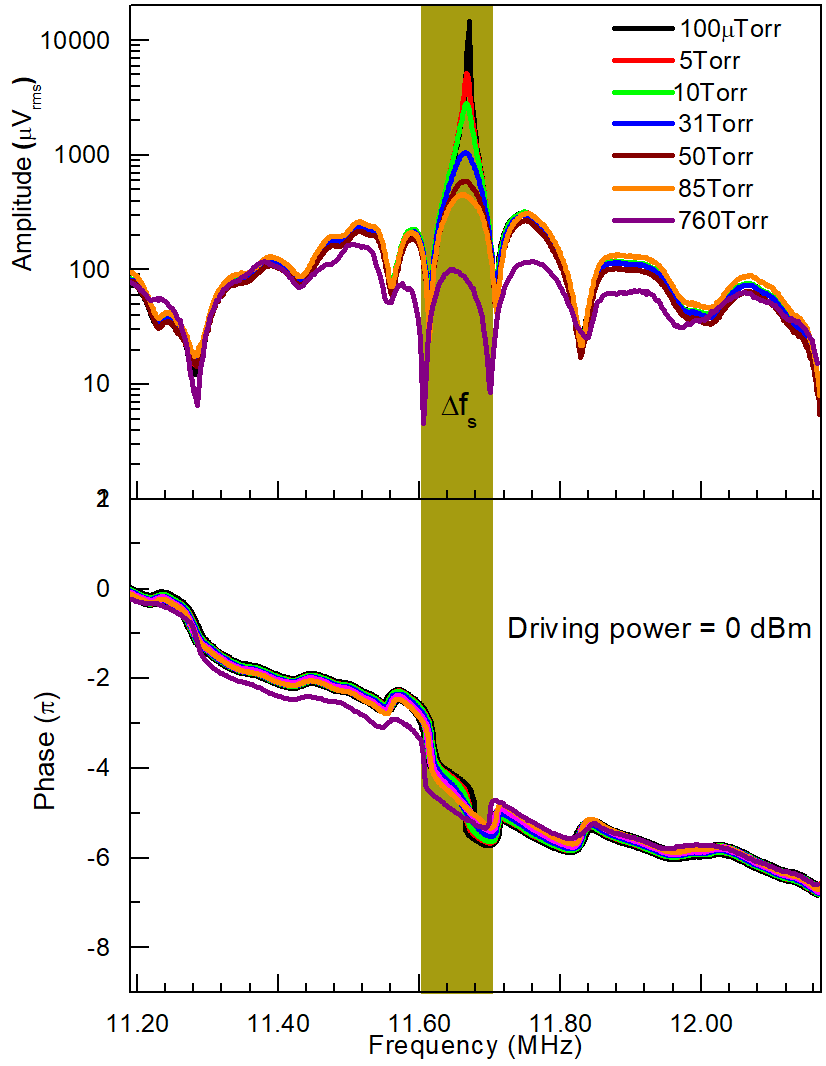}
\vspace{0.5 em}
\includegraphics{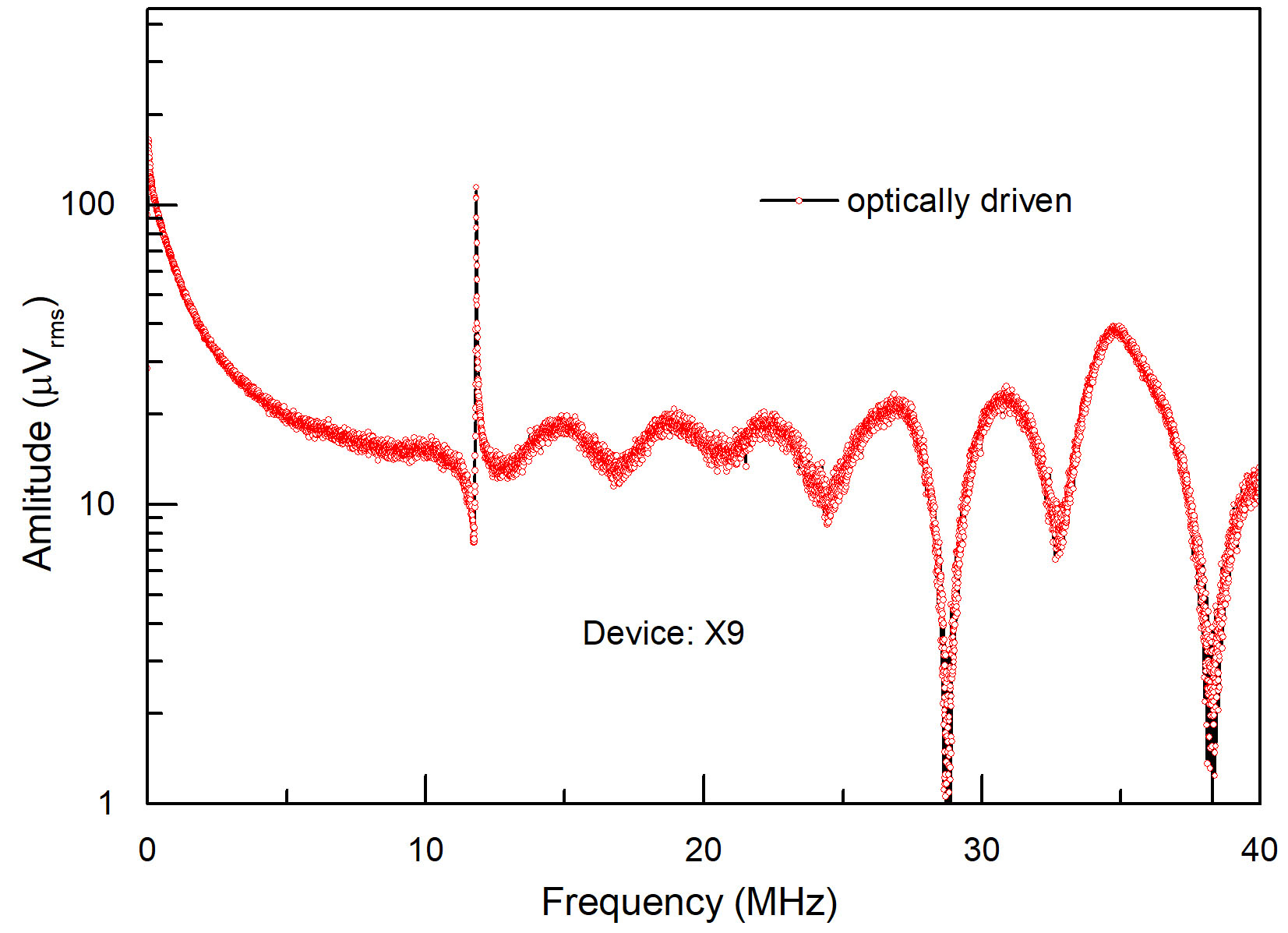}
\caption[Figure S2 short description]{\textbf{a) Evolution of mechanical resonance shape of a similar device to that described in main text by piezoactuation from high vacuum to atmospheric pressure.} A forest of acoustic peaks ($\Delta f_{s}$s) can be seen either side of the resonance peak at all pressures. The shaded area is the characteristic frequency span, $\Delta f_{s}$  due to acoustic wave interference within which mechanical resonance can be seen. Mechanical resonance is showing a strong dependence on damping in contrast to surface motion. Phase evolutions of same experiments are shown at the bottom.
 \textbf{b) Measurements of optomechanically driven responses at 15 Torr.} Surface acoustic wave interference is absent.\label{s:fig:bulkacoustic}}
\end{figure}

Figure \ref{s:fig:bulkacoustic}a shows amplitude and phase response of a single NEMS device where the frequency span crosses 8 or 9 bulk acoustic peaks.  The driving power is kept constant at 0 dBm as scans are taken at differing pressures (and damping conditions).  Up to about 50 Torr, the background region outside of the span $\Delta f_s$ is almost identical.  The pressure changes have essentially no effect on bulk acoustic resonances, as would be expected.  The signal to background ratio of the NEMS resonance peaks (against this bulk acoustic background) range from about 60x to 3x and the NEMS peaks are easily identifiable.  For 85 and 760 Torr responses, the NEMS resonance widths are wider than $\Delta f_{s}$, and the NEMS amplitude contribution to the signals is comparable to the bulk acoustic resonance contributions.  Thus, extra care needs to be taken when identifying NEMS resonance peaks at highest damping, for example, by tracking the peak from vacuum to atmosphere, to properly identify the appropriate locking frequency range (in this case, within the $\Delta f_{s}$ span).
To fully confirm the nature of the acoustic wave interference during piezodrive, we measured the same device with optomechanical drive and the comparison is shown in Fig. \ref{s:fig:bulkacoustic}b for a wide span.  The optical drive response does not see the forest of bulk acoustic resonances, as expected.  The optical drive has its own background due to imperfect filter extinction of the drive laser at the photoreceiver \cite{Sauer2017}, with its own 4 MHz interference pattern, but this is  irrelevant for the present work.

\subsubsection{Squeeze film effects}

There is a small gap (140 nm) between our nanomechanical devices and the waveguides in the optical ring resonator. This geometry could indicate squeeze film effects, wherein the air in the gap can act to increase the effective stiffness of the nanomechanical beam and hence affect its dynamic behaviour. Using the dimensionless squeeze number \cite{Bao2007, Blech1983} for strip plates we can determine whether viscous or spring effects are dominant. The squeeze number is defined as
\begin{equation}
\sigma =   (12\mu L^2 \omega)/(P_a {h_a}^2 )
\end{equation}
where $\sigma$ is the dimensionless squeeze number, $\mu$ is the dynamic viscosity ($Nsm^{-2}$) of the medium, $L$ is the characteristic length scale (here it is the width of the nanomechanical beam, 220 nm), $\omega$ is the angular frequency of the nanomechanical beam, $P_a$ is the pressure of the medium, and $h_a$ is the gap between the beam and the photonic waveguide. In practice, $\sigma <1$ signifies a regime when squeeze film spring effects are not important and that viscous damping effects are dominant. Using the values for our primary device, we calculate a squeeze number of 0.4, which implies viscous damping is the dominant effect. It is not important to our general analysis what precisely causes the damping at higher pressures (whether it be pure viscous air damping or squeeze film air pot damping), therefore, we conclude that further squeeze film analysis is unnecessary.

\subsection{Lock-in amplifier and PLL details}

A phase-locked loop (PLL) is essentially a feedback control system which locks the phase and frequency output of a low noise oscillator to the phase and frequency of an input signal.  In a sensing context, it can be used to stabilize and track the resonance frequency of the input signal, which carries the sensed information in its resonance frequency. Extensive applications of PLL for tracking nanomechanical vibration can be found in Ref. [\!\citenum{ giessibl2003advances}] and the references therein for atomic force microscopy.  Roukes' group pioneered analog PLL use in NEMS for mass sensing \cite{Yang2006}.  Recently, Olcum \textit{et al.} \cite{Olcum2015} gave a very detailed discussion of loop dynamics during the use of a closed loop PLL for measuring stability and mass sensitivity. We use a PLL in closed loop to track frequency shifts for the purposes of determining stability (such as for Allan deviation measurements) as well as for tracking frequency shifts caused by mass adsorbants \cite{Venkatasubramanian2016} or due to temperature change (\textit{cf.} SI section 1.7). We use open loop measurements for verification of presence or absence of intrinsic frequency fluctuation noise (as in SI section 2.5).

Figure \ref{s:fig:Spll} describes our PLL circuit, which basically takes advantage of the built in functionality of the Zurich Instruments HF2LI.  The NEMS as the device under test is the frequency determining element in the circuit, controlling the NCO frequency in the Zurich instrument via PID feedback.  The feedback controller and the PID parameters control the PLL bandwidth via the PID gains, creating a transfer function for the error signal.  Fluctuations on a faster time scale than the corner frequency of the transfer function start to become filtered out.  Thus, sampling times $\tau$ shorter than the inverse of the PLL bandwidth are generally not reported.  The demodulator portion of the circuit measures the instantaneous frequency and phase of the incoming signal.  It has a demodulation bandwidth set by its low pass filter that is kept at $8$ times the PLL bandwidth for stability reasons.  For purposes of noise measurement, the demodulation bandwidth is what sets the noise measurement bandwidth $\Delta f$ and the high frequency integration cutoff $f_{\mathrm{H}}$ discussed in SI section 2.

\begin{figure}[ht]
    \centering
    \includegraphics[scale=1]{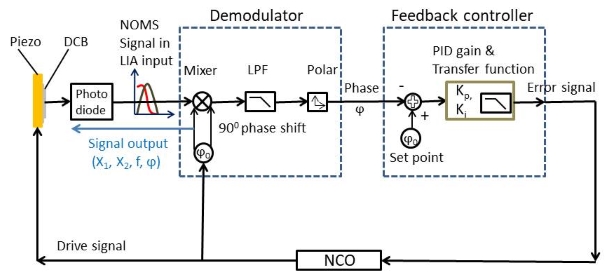}
    \caption{\textbf{A dual-phase demodulator, a controller, and an NCO are three essential building blocks of a phase locked loop configuration inside the Zurich instrument HF2 lock-in amplifier.} These three are combined to form a negative feedback loop. Within the loop, the phase detector (mixer) detects the phase difference between the incoming NOMS signal and the reference. Depending on PI gain (set by the bandwidth, D parameter is not in use) the controller regulates the NCO to achieve a vanishing phase difference, which means that the NCO frequency always adapts the NEMS frequency at a constant SNR by maintaining a -90 phase between the DUT and NEMS. Thus the lock-in output, i.e., the reference always follows the NEMS frequency depending on phase error controlled by the feedback and overlooks any error due to amplitude fluctuations.}
    \label{s:fig:Spll}
  \end{figure}

The PID parameters are automatically calculated by the lockin "advisor" software based on mechanical $Q$, center frequency, desired PLL bandwidth, locking range, and phase setpoint.  We have primarily chosen 500 Hz (with a 24 dB/oct filter) as the PLL loop bandwidth for data presented.  The advisor computes through a numerically optimized algorithm of loop dynamics to generate a set of feedback gain parameter which tries to match the target bandwidth in its simulated first-order transfer function. Figure \ref{s:fig:SPLLbode} shows a representative bode plot of an advisor simulated transfer function for 500 Hz PLL BW which has a $ 3\; \textup{dB}$ roll-off at 500 Hz and is a typical example of PLL transfer function.

 \begin{figure}[ht]
    \centering
    \includegraphics[scale=1]{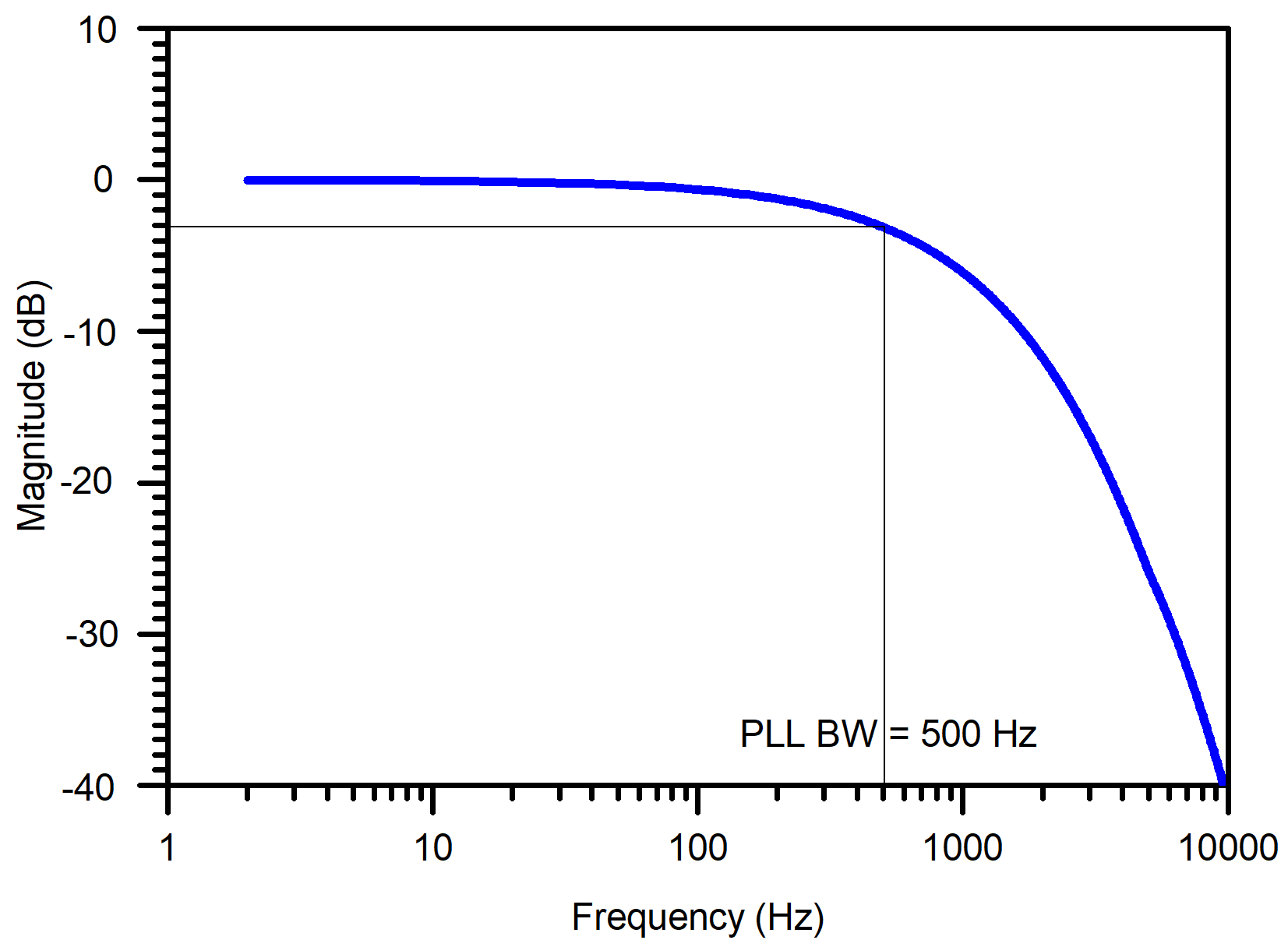}
    \caption{\textbf{A representative PLL transfer function obtained from Zurich instrument HF2.} Target BW is at $-3$ dB point in the bode plot. In case of any mismatch between set resonance parameters, a target bandwidth, and numerical modeling advisor fails to produce such bode plot with warning indications. It automatically adjusts the demodulation bandwidth to value eight times PLLBW to avoid being  limited by the demodulation speed.}
    \label{s:fig:SPLLbode}
  \end{figure}

Our present understanding of one advantage of a tight PLL over a self-oscillating circuit is the following.  The latter allows random walk phase noise (e.g. coming from thermomechanical noise phase walking) that is not present when using a stable external source. The price paid is that the phase noise of the external source is injected into the system.  In the present case, that noise is negligible in comparison to the measured Allan deviations (the Rb time base source is quoted with a $5 \times 10^{-11}$ stability at 1 s, corresponding to $\leq 0.2 \times 10^{-9}$ at $1$ ms).  We believe that the elimination of this source of random walk phase noise may play an important role in exposing the flatband nature of thermomechanical phase noise values in our experiment, and ultimately to our Allan deviation measurements agreeing closely to Eqn. 10.  

\subsection{Temperature measurement calibration procedure}
A Nano-optomechanical system (NOMS) includes a high-quality optical cavity or a microring resonator coupled to a mechanical resonator in nanometric dimensions. In the current work, the mechanical element is a double clamped beam (DCB). Both the optical ring and NEMS in the integrated NOMS structure are susceptible to environmental fluctuations, and consequently, both may be used as temperature sensors. A small temperature change on the device surface changes simultaneously the resonance wavelength, $ \lambda _{0}$ of the optical ring and the resonance frequency, $f _{0}$ of the NEMS. $\lambda _{0}$ of optical spectra changes with  temperature mainly due to the thermo-optic effect of silicon \cite{Rouger2010}. Quantities such as elastic modulus and thermal expansion coefficient of silicon determine the resonance frequency of NEMS which depends on temperature strongly \cite{Zhang2013,melamud2007temperature,Inomata2016}. By using a PID controlled heater, we can modify the chip surface temperature and test both the NOMS and ring as thermometers, effectively calibrating them against each other.

\subsubsection{Microring thermometry}
Details of device configuration and principle have been described in detail \cite{Sauer2014}. A change in temperature $\Delta T$ will shift ring properties via thermal expansion of silicon and oxide and via thermo-optic coefficient (TOC) of Si, $\alpha_{\textup{n}_{\textup{Si}}}=2\times 10^{-4}\;\textup{ K}^{-1}$. The latter is the dominant effect.  This will give a temperature responsivity $S_{\lambda ,\textup{T}} \equiv \frac{d\lambda_0}{dT}$ that can be theoretically approximated by
\begin{equation}
    S_{\lambda , \textup{T}} \cong  \frac{\lambda_0 \alpha_{\textup{n}_{\textup{Si}}}}{n_{\textup{Si}}}
\end{equation}
which gives approximately $80$ pm/K for 1550 nm light \cite{Rouger2010}.

In our system, we use the probe sitting on the side of the optical resonance to transduce $\Delta \lambda$ due to temperature change into $\Delta T_r$, the change in transmission, through the slope responsivity, $\Re_{\lambda} \equiv dT_\textup{r}/d\lambda$.  This gives, finally
 \begin{equation}
\Delta T_{\textup{ring}}=\frac{\Delta \lambda_0}{S_{\lambda , \textup{T}}}=\frac{\Delta T_{\textup{r}}}{S_{\lambda , \textup{T}} \Re_{\lambda}}
  \end{equation}

Both $S_{\lambda , \textup{T}}$ and $\Re_{\lambda}$ can be measured experimentally. $S_{\lambda , \textup{T}}$ is calibrated by setting known temperature changes into the PID temperature controller and extracting $\Delta \lambda_0$ values from static temperature wavelength sweeps. $\Re_{\lambda}$ is observed directly from wavelength sweep slope at the probe point.
  
\subsubsection{NOMS thermometry}
 The fundamental flexural mode eigenfrequency of a straight doubly clamped beam (without residual tension) made of homogeneous material is \cite{Cleland2002}
 \begin{equation}
  f_{0}= 1.027 \frac{t}{l^{2}}\sqrt{\frac{E}{\rho }}
  \label{s:eq:fDCB}
  \end{equation}
  where, \textit{t} and \textit{l} are the thickness and the length of the beam, \textit{E} and $\rho $ are the elastic moduli and density of the material.  For a beam with residual tension such as compressive stress $\sigma_i$, the frequency modifies to \cite{jun2006electrothermal}
  \begin{equation}
 f_{\sigma _{\textup{i}}}= f_{0}\sqrt{1-\frac{0.295\sigma _{\textup{i}}l^{2}}{Et^{2}}}
  \label{s:eq:fDCBcT}
  \end{equation}  
 
All quantities on the R.H.S. of equations \ref{s:eq:fDCB} and \ref{s:eq:fDCBcT} change with temperature. As a consequence, the resonance frequency of nanomechanical resonators strongly depends on temperature. This \textit{f-T} relationship is referred to as the temperature coefficient of resonant frequency, \textit{TCRF}which is the ratio of temperature sensitivity ($S_{f,T}=\frac{df}{dT}$) to its resonance frequency, $f_{0}$. i.e.
 \begin{equation}
TCRF= \frac{1}{f_{0}}\frac{df}{dT}= \frac{1}{f_{0}}S_{\textup{f,T}}
  \end{equation}
  $S_{\textup{f, T}}$ can be measured experimentally by identifying $f_{0}$ from thermomechanical noise spectra taken at different set temperatures. 
Thus measured temperature from temperature induced frequency shift of PLL data can be found as
 \begin{equation}
\Delta T_{\textup{NEMS}}= \frac{\Delta f}{S_{\textup{f, T}}}
  \end{equation} 
  
\subsubsection{Static temperature measurement}
 
 An example of the calibration of ring ($S_{\lambda ,\textup{T}}$) and NEMS ($S_{f ,\textup{T}}$) temperature responsivity is given in Fig. \ref{s:fig:staticT} for 3 Torr pressure.  Increasing temperature causes a red shift in optical ring wavelength and a decrease in the resonance frequency.  Measurements of temperature sensitivities (by both ring and NOMS) at different pressures are shown in Table 1.  The slight drop in sensitivity with increasing pressure may be due to the surface not fully reaching the temperature change set by the PID and measured by the Pt RTD at the copper base due to increased heat transfer coefficient of the higher pressure air.  Both surface sensors show a consistent measurement.
 \begin{figure}[ht]
\centering
\includegraphics{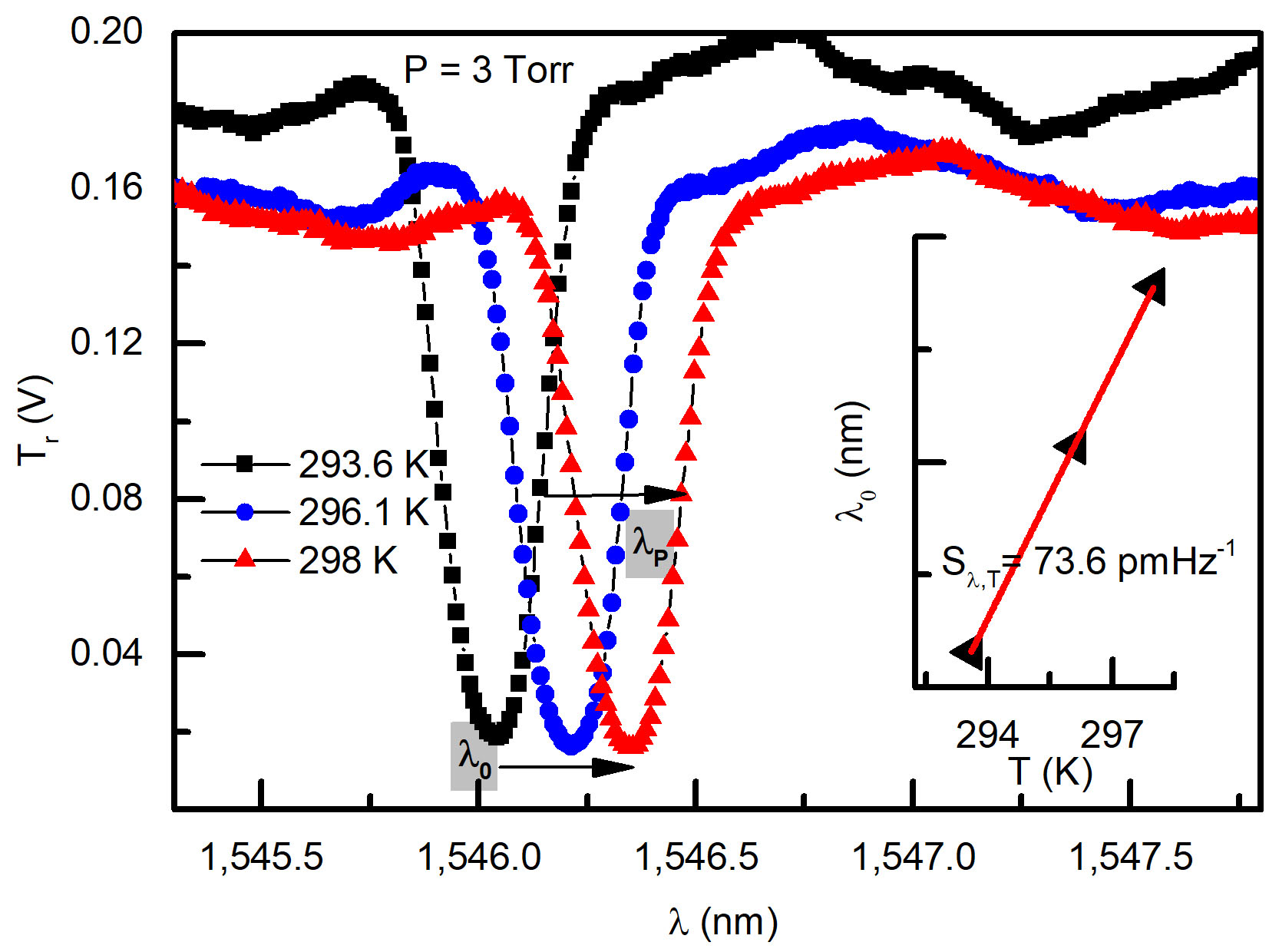}
\vspace{0.5 em}
\includegraphics{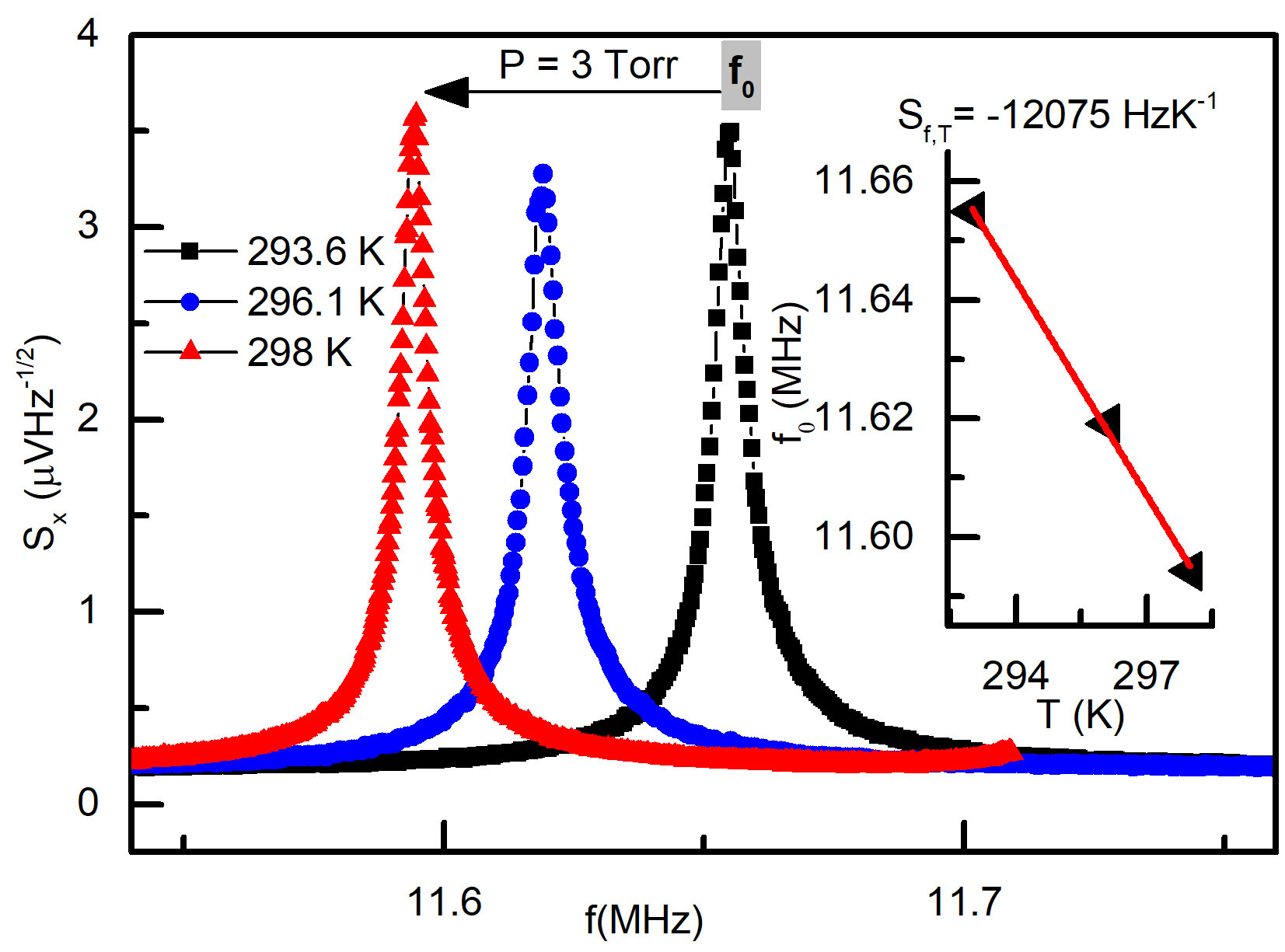}
\caption[Figure S9 short description]{\textbf{Representative plots for determining $S_{\lambda ,\textup{T}}$ and $S_{ \textup{f,T}}$.}  These are found from the linear temperature dependence of resonance wavelength of the optical ring (left) and the resonance frequency of the NOMS (right). \label{s:fig:staticT}}
\end{figure}
The temperature sensitivity of the optical ring, at around 70 to 80 pm/K, is consistent with the literature \cite{Rouger2010}. The TCRF of the NOMS in this device ranges from -1050 to -1270 ppm/K, which is an order of magnitude larger than expected from materials properties alone.  In another chip, the values ranged from -140 to -340 ppm/K. This discrepancy can be explained by the residual tension within the NEMS doubly clamped beams\cite{}. Changes to the temperature can have a much larger effect on the resonant frequency through modifying this tension than the material properties.
\begin{table}[h]
\centering
\caption{Measured ring and NOMS temperature sensitivity at different pressures}
\label{my-label}
\begin{tabular}{|c|c|c|c|}
\hline
P(Torr) & \begin{tabular}[c]{@{}c@{}}Ring sensitivity\\ $S_{\lambda ,T}=\frac{d\lambda }{dT} [pmK^{-1}]$\end{tabular} & \begin{tabular}[c]{@{}c@{}}NEMS sensitivity\\ $S_{f,T}=\frac{df }{dT} [(kHz)K^{-1}]$\end{tabular} & \begin{tabular}[c]{@{}c@{}}TCRF\\ $\frac{1}{f_{0}}\frac{df}{dT}\times 10^{6}[ppm]$\end{tabular} \\ \hline
100$\mu$ & 81 $\pm$ 5 & -14.7 $\pm$ 0.2 & -1269 $\pm$ 19 \\ \hline
3 & 73.6 $\pm$ 0.3 & -12.1 $\pm$ 0.4 & -1041 $\pm$ 35 \\ \hline
61 & 76.1 $\pm$ 1.4 & -13.4 $\pm$ 0.2 & -1156 $\pm$ 13 \\ \hline
760 & 70.5 $\pm$ 2.5 & -12.1 $\pm$ 0.6 & -1046 $\pm$ 54 \\ \hline
\end{tabular}
\end{table}
\subsubsection{Origin of higher TCRF in NEMS doubly clamped beams}
From the beam geometry, $\textup{t} =160$ nm, $l= 9.75$ $\mu \textup{m}$   and materials values, $E=170$ GPa, $\rho=2330\;  \textup{kgm}^{-3}$the expected resonance frequency of the device from equation \ref{s:eq:fDCB} can be found as $14.8$ MHz.  Measured frequency is quite different at $11.8$ MHz.  This is likely an indication of residual compressive stress. Rearranging equation \ref{s:eq:fDCBcT} we have
 \begin{equation}
  \sigma _{\textup{i}}=3.4E \frac{f_{0}^{2}-f_{\sigma_\textup{i}}^{2}}{f_{0}^{2}}\frac{t^{2}}{l^{2}} 
  \end{equation} 
which allows estimating the residual compressive stress as $57$ MPa. If the beam is heated, the compressive stress will change, ultimately changing the frequency.  The total stress can be set as an initial stress plus a thermal induced stress. 
 \begin{equation}
 \sigma =\sigma _{\textup{i}}+\sigma _{\textup{t}}= \sigma _{\textup{i}}-\alpha _{\textup{l}}E\Delta T
   \end{equation}  
where $\alpha_\textup{l}$ is the thermal expansion coefficient. This gives a temperature coefficient of thermal stress due to initial strain 
\begin{equation}
 \alpha _{\sigma }=\frac{1}{\sigma _{\textup{i}}}\frac{d\sigma }{dT}\approx \frac{\sigma -\sigma _{\textup{i}}}{\sigma _{\textup{i}}dT}=\frac{-\alpha_{\textup{l}} E}{\sigma _{\textup{i}}} 
   \end{equation} 
After substituting in values we find, $\alpha _{\sigma }= -7780 \;  \textup{ppmK}^{-1}$.
Inomata \textit{etal.}\cite{Inomata2016} deduce the analytical expression for temperature coefficient of resonance frequency, \textit{TCRF} for a stressed double clamped beam as follows
 \begin{equation}
TCRF= \frac{1}{2}\alpha _{\textup{E}}-\alpha _{\textup{l}}-\frac{1}{2}\alpha _{\rho }+\frac{1}{2}\frac{0.295\epsilon \frac{l^{2}}{t^{2}}}{1+0.295\epsilon \frac{l^{2}}{t^{2}}}\alpha _{\sigma } 
 \label{s:eq:TCRFtension}
 \end{equation}  
where $\alpha_\textup{E}$, $\alpha_{\rho}$, and $\alpha _{\sigma }$ are the temperature coefficients of Young's modulus, density, and thermal stress, respectively. Also, $\alpha_\textup{l}$ is the thermal expansion coefficient, $\varepsilon$ is the initial strain (calculated value of $334\times 10^{-6}$ for the device in the current work), and $l$ and $t$ are the device length and flexure-direction thickness, respectively.
 Plugging values for our device into equation \ref{s:eq:TCRFtension} we find a $ TCRF =  -1078\;\textup{ppmK}^{-1}$ which is  in good agreement  with our experimental results (around 
$ 1100-1200\;\textup{ppmK}^{-1}$) displayed in Table I.

\subsubsection{Dynamic temperature measurement and discussion}

An example of dynamic temperature measurements is given in Fig. \ref{s:fig:Tonoff} measured at $3$ Torr for $1$ kHz PLL bandwidth.
\begin{figure}[h]
    \centering
    \includegraphics[scale=1]{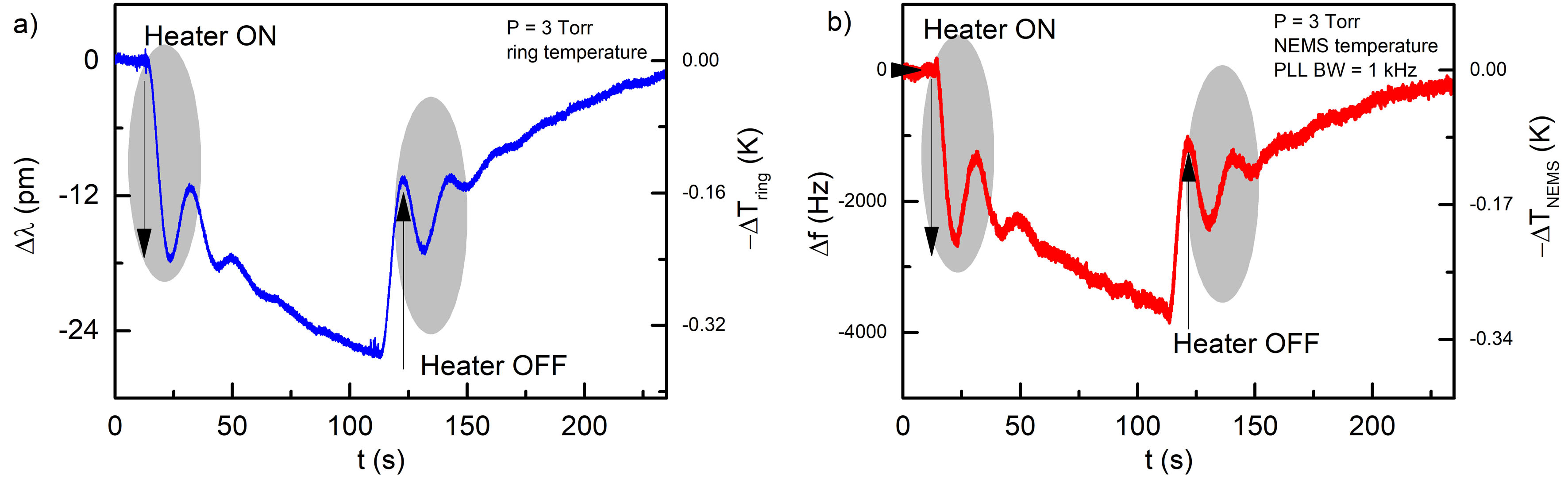}
    \caption{\textbf{An illustration of the change in a) wavelength and b) resonance frequency with time during step changes on and off in temperautre.} A PID controlled heater steps from $298$ K to $298.3$ K followed by a cooling step back to $298$ K. The ring data in figure (a) and NOMS data in figure (b) were measured simultaneously during the same heating and cooling cycle. The largest temperature difference measured (just when the heater was turned off to return at $298$ K) by the ring is $\approx 0.35$ K and by the NEMS is $\approx 0.32$ K.  The shaded areas highlight overshooot of the PID temperature control loop.}
    \label{s:fig:Tonoff}
  \end{figure}
Both temperature sensors shows similar transient response with temperature change. The most probable source of error in $ S_{f,T}$ measurements is the inability of establishing surface temperature equilibrium within approximately 15 minutes waiting span used to move from one set temperature to another. The values of $S_{f,T}$ and $S_{\lambda,t}$ therefore may be slightly overestimated by the static method, in comparison to Fig \ref{s:fig:Tonoff}. We confirmed this qualitatively in later measurements using a separate temperature sensor glued at the top of the chip surface.  

\section{Frequency Stability}
\subsection{Robins' phase noise analysis}

The white force noise, due to thermal energy which is normalized to give $1/2$ $\kb T$ after the integration over a mechanical mode resonance as stated by the equipartition theorem, is defined as follows:
\begin{equation} 
S_\mathrm{F}^\mathrm{th} (\Omega) = 4 M \Gamma \kb T.
\end{equation}
This force noise is shaped into a Lorentzian displacement noise by the mechanical susceptibility, $\chi$, of the mechanical resonator (i.e. the mechanical transfer function):
\begin{equation} 
S_\mathrm{x}^\mathrm{th} = \chi^2 S_\mathrm{F}^\mathrm{th},
\end{equation}
where
\begin{equation} 
\chi(\Omega) = \frac{1}{M ( \Omega_0^2 - \Omega^2 - i \Gamma \Omega)}.
\end{equation}
Above, $\Gamma = \Omega / Q$ where $Q$ is the mechanical quality factor, $\Gamma$ is the resonant linewidth (i.e. damping), and $M$ is the effective mass of the mechanical resonant mode. 

\begin{figure}[h]
\centering
\includegraphics[scale=1.5]{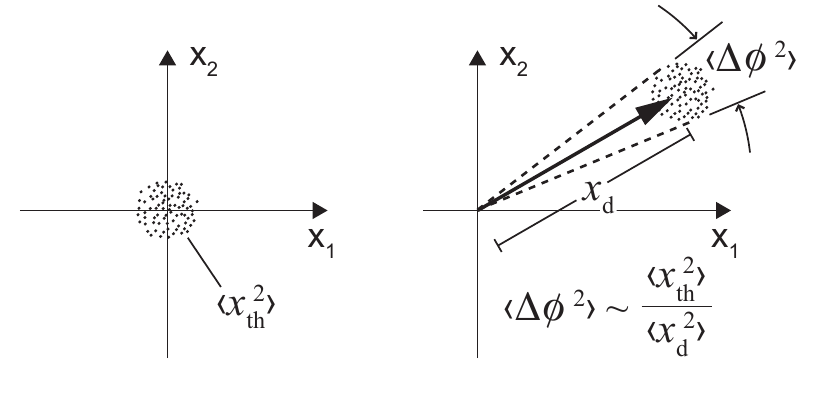}
\caption[]{\textbf{Conceptual sketch of the phase noise definition.} \label{s:fig:phasenoise}}
\end{figure}

From Figure~\ref{s:fig:phasenoise}, the phase noise is taken as
\begin{equation} \label{s:eq:phasenoise}
S_\phi^\mathrm{th} (\Omega) = \frac{1}{2} \frac{S_\mathrm{x}^\mathrm{th}}{\langle x_\mathrm{d}^2 \rangle}.
\end{equation}
The factor of 1/2 comes from the property that 1/2 the noise will be in the amplitude quadrature and 1/2 will be in the phase quadrature. Also from the figure, the average squared thermal amplitude is defined as
\begin{equation} 
\langle x_\mathrm{th}^2 \rangle = \frac{\kb T}{M \Omega_0^2}.
\end{equation}

If we define $f$ as the offset from the carrier frequency such that $\omega = 2 \pi f = \Omega - \Omega_0$ we find that
\begin{equation} 
\left | \chi (\Omega) \right |^2 = \frac{1}{(2 \Omega_0 \omega + \omega^2)^2 + \Gamma^2 \Omega^2}  \frac{1}{M^2},
\end{equation}
thus,
\begin{equation} 
S_\phi^\mathrm{th} (\Omega) = \frac{1}{2} \frac{1}{(2 \Omega_0 \omega + \omega^2)^2 + \Gamma^2 \Omega^2}  \frac{4 \Gamma  \kb T}{M \langle x_\mathrm{d}^2 \rangle}.
\end{equation}

\begin{figure}[h]
\centering
\includegraphics{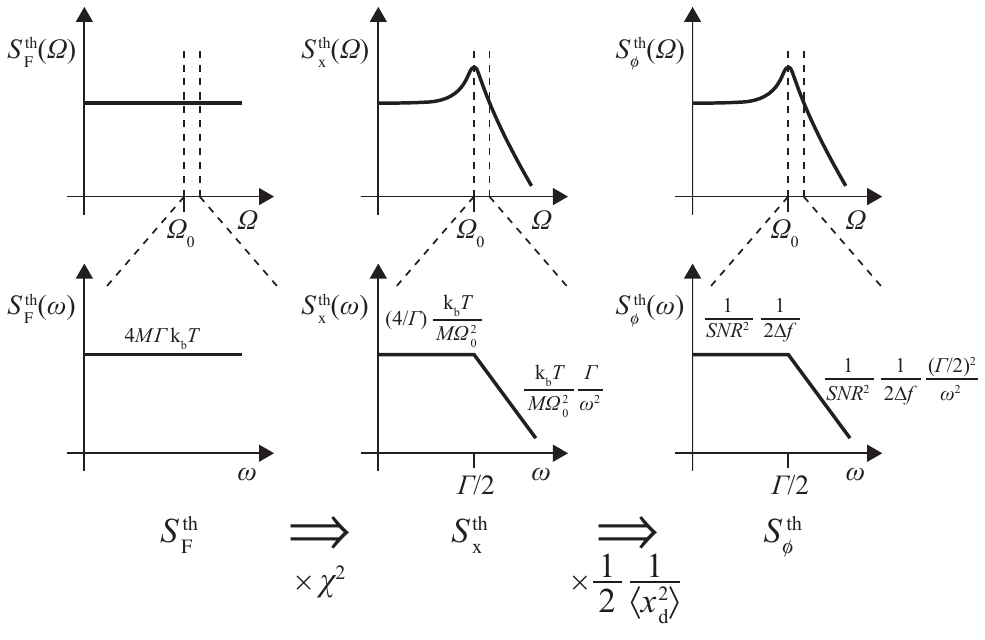}
\caption[]{Conceptual diagram of the force noise translating to phase noise. \label{s:fig:force_displ_phase}}
\end{figure}

Normally, the following assumptions are made: (1) $\omega <\!\!< \Omega_0$, and (2) $\omega >\!\!> \Gamma$. These simplify the derivation to result in \cite{Cleland2002, Albrecht1991}
\begin{equation} \label{s:eq:phasenoisesimple}
S_\phi^\mathrm{th} (\omega) \approx \frac{\langle x_\mathrm{th}^2 \rangle}{\langle x_\mathrm{d}^2 \rangle}  \frac{\Gamma/2}{\omega^2}.
\end{equation}
However, for moderate and higher damping, and for frequencies close to the carrier, condition (2) no longer holds. Simplifying using only condition (1) we obtain
\begin{equation} \label{s:eq:phasenoisesimple2}
\begin{split}
S_\phi^\mathrm{th} (\omega) & \approx \frac{1}{(2\Omega_0 \omega)^2 +\Gamma^2\Omega_0^2} \; \frac{2\Gamma\kb T}{M \langle x_\mathrm{d}^2 \rangle} \\
& \approx \frac{1}{(2\Omega_o)^2} \; \frac{1}{\omega^2 + (\Gamma/2)^2} \; \frac{2\Gamma\kb T}{M \langle x_\mathrm{d}^2 \rangle} \\
S_\phi^\mathrm{th} (\omega) & \approx \frac{\langle x_\mathrm{th}^2 \rangle}{\langle x_\mathrm{d}^2 \rangle}  \; \frac{\Gamma/2}{\omega^2 + (\Gamma/2)^2}.
\end{split}
\end{equation}

Next, if we define $a_\mathrm{th}^2 \equiv S_\mathrm{x}^\mathrm{th} (\Omega_o) \Delta f$, where $\Delta f$ is the measurement bandwidth of the $S_\mathrm{x}^\mathrm{th}$ quantity
\begin{equation} \label{s:eq:ath}
\begin{split}
a_\mathrm{th}^2 & = \frac{4 \kb T}{M \Omega_0^2 \Gamma} \Delta f \\
& = \frac{\langle x_\mathrm{th}^2 \rangle}{\Gamma/4} \Delta f
\end{split}
\end{equation}
we can then define
\begin{equation} \label{s:eq:SNRsq}
(\SNR)^2 \equiv \frac{\langle x_\mathrm{d}^2 \rangle}{a_\mathrm{th}^2} = \frac{\langle x_\mathrm{d}^2\rangle \Gamma/4}{\langle x_\mathrm{th}^2 \rangle \Delta f}.
\end{equation}
Therefore, $S_\phi^\mathrm{th}(\omega)$ can finally be written as:
\begin{equation} \label{s:eq:phasenoiseSNR}
S_\phi^\mathrm{th}(\omega) = \frac{1}{(\SNR)^2} \; \frac{1}{2 \Delta f} \; \frac{(\Gamma /2)^2}{\omega^2 + (\Gamma/2)^2}
\end{equation}

The shape of $S_\phi^\mathrm{th}(\omega)$ is thus a low pass filter with a knee at $\omega = \Gamma /2$; it can be approximated as a constant value near the carrier frequency and as a $1/\omega^2$ function far from the carrier frequency:
\begin{align} 
S_{\phi\mathrm{,near}}^\mathrm{th}(\omega) & \sim \frac{1}{(\SNR) ^2} \frac{1}{2 \Delta f} &\omega <\!\!< (\Gamma/2) \label{s:eq:phasenoisenear} \\
S_{\phi\mathrm{,far}}^\mathrm{th}(\omega) & \sim \frac{1}{(\SNR)^2} \frac{1}{2 \Delta f} \frac{(\Gamma /2)^2}{\omega^2}.  &\omega >\!\!> (\Gamma/2) \label{s:eq:phasenoisefar}
\end{align}

\subsection{Definition of Allan Deviation} \label{allandevdefinition}

Allan deviation, $\sigma_y$, is defined as the square root of the Allan variance, $\sigma_y^2$, 
\begin{equation} \label{s:eq:avar}
\sigma_y (\tau) = \sqrt{\sigma_y^2(\tau)} = \sqrt{\frac{1}{2}\left \langle (\bar{y}_{n+1} - \bar{y}_n)^2 \right \rangle}.
\end{equation}
$\tau$ is the observation period and $\bar{y}_n$ is the $n$th fractional frequency average over the observation time. The relationship between close-in frequency or phase noise and Allan variance (worked out primarily at NIST in the 1960s and 70s~\cite{Barnes1971}) integrates the noise with a transfer function $H(f,\tau)$ as below
\begin{equation} \label{s:eq:NISTavar}
\sigma_y^2(\tau) = 2 \int_0^{f_H} S_y(f) \frac{\sin^4(\pi \tau f)}{(\pi \tau f)^2} \mathrm{d}f
\end{equation}
where
\begin{equation*}
S_y(f)  \equiv \frac{f^2}{\nu^2} S_\phi (f), \qquad \mathrm{in~which}~f=\omega/(2 \pi) ~\mathrm{and}~\nu = \Omega/(2 \pi).
\end{equation*}
and the transfer function is
\begin{equation*}
H(f,\tau) \equiv \frac{\sin^4(\pi \tau f)}{(\pi \tau f)^2}
\end{equation*}

For $S_y(f)$ exhibiting power law behaviour there are known power law solutions to equation \ref{s:eq:NISTavar}:
\begin{equation}
\sigma_y^2 (\tau)= \left\{\begin{array}{ll}
 A f^2 S_y(f)\tau^1  & \mathrm{for }~S_y(f) \sim f^{-2}; (S_\phi (f)\sim f^{-4}) \\ 
 B f^1 S_y(f)\tau^0  & \mathrm{for }~S_y(f) \sim f^{-1}; (S_\phi (f)\sim f^{-3}) \\
 C f^0 S_y(f)\tau^{-1}& \mathrm{for }~S_y(f) \sim f^{0}; (S_\phi (f)\sim f^{-2}) \\ 
 D f^{-1} S_y(f)\tau^{-2} & \mathrm{for }~S_y(f) \sim f^{1}; (S_\phi (f)\sim f^{-1}) \\ 
 E f^{-2} S_y(f)\tau^{-3 }& \mathrm{for }~S_y(f) \sim f^{2}; (S_\phi (f)\sim f^{0}) \\
\end{array}\right.
\end{equation}
where
\begin{equation*}
\begin{split}
A &= 4 \pi^2/6 \\
B &= 2 \ln 2 \\
C &= 1/2 \\
D &= 1.038 + 3 \ln (2\pi f_H \tau_0) / (4 \pi^2) \\
E &= 3 f_H / (4 \pi^2)
\end{split}
\end{equation*}
Then, following from equations \ref{s:eq:phasenoisenear} and \ref{s:eq:phasenoisefar}
\begin{align} 
S_{y\mathrm{,near}}(f) & \cong \frac{1}{(\SNR) ^2} \frac{1}{2 \Delta f} \frac{1}{\nu^2}f^2 & f <\!\!< \Gamma/(4\pi) \label{s:eq:freqnoisenear} \\
S_{y\mathrm{,far}}(f) & \cong \frac{1}{(\SNR)^2} \frac{1}{2 \Delta f} \frac{1}{(2\pi)^2 \nu^2}\left ( \frac{\Gamma}{2} \right )^2.  &f >\!\!> \Gamma/(4\pi) \label{s:eq:freqnoisefar}
\end{align}
This implies that, assuming $\Delta f = f_H$ is the measurement bandwidth,
\begin{align}
\sigma_{\mathrm{fb}}(\tau) \equiv \sigma_{y\mathrm{,near}}(\tau) & = \left ( \frac{3}{2}\right )^{1/2}\frac{1}{\SNR}\frac{1}{\Omega}\frac{1}{\tau} \label{s:eq:allandevnear} \\
\sigma_{\mathrm{R}}(\tau) \equiv \sigma_{y\mathrm{,far}}(\tau) & = \frac{1}{4}\frac{1}{\SNR}\frac{\Gamma}{\Omega}\frac{1}{(\tau \Delta f)^{1/2}} = \frac{1}{4}\frac{1}{\SNR}\frac{1}{Q}\frac{1}{(\tau \Delta f)^{1/2}} \label{s:eq:allandevfar}.
\end{align}
Equation \ref{s:eq:allandevfar} is essentially Robins' formula (denoted with subscript "R").  Equation \ref{s:eq:allandevnear} applies when measurement bandwidth is less than the linewidth, a situation which we will refer to as the flatband regime and denote with the subscript "fb".  From these equations it can be seen that for the situation where $\SNR \propto 1/Q^{1/2}$ (a usual case in AFM), $\sigma_{\mathrm{R}} \sim Q^{-1/2}$ as expected, but $\sigma_{\mathrm{fb}} \sim Q^{+1/2}$. For situations where $\SNR \propto 1/Q$, such as when accessing full dynamic range, $\sigma_{\mathrm{R}} \sim Q^{0}$ (no $Q$ dependence) and $\sigma_{\mathrm{fb}} \sim Q^{1}$ (better stability for lower $Q$). These two situations are considered in Figs 4 and 5.

\subsection{Allan deviation integrations}

This section demonstrates some of the integrations used to arrive at the results in Section~\ref{allandevdefinition}. We will first start at the basic equation (which is equivalent to equation~\ref{s:eq:NISTavar})
\begin{equation}
\sigma^2(\tau) = \frac{1}{\pi}\left(\frac{2}{\Omega \tau}\right)^2 \int_{0}^{f_H 2 \pi} \mathrm{d}\omega S_{\phi}(\omega) \sin^4\left(\frac{\omega \tau}{2}\right) \label{s:eq:allanvarbasic}.
\end{equation}
Using $\sin^4\left(\frac{\omega \tau}{2}\right) \equiv \frac{1}{8} \left [ 3 - 4 \cos (\omega \tau) + \cos(2\omega \tau) \right ]$, along with letting $2\pi f_H \rightarrow \infty$ and using these additional integral identities,
\begin{equation}
\int_0^\infty \frac{\mathrm{d}x}{x^2 + b^2} =\left.  \frac{1}{b}\arctan \left (\frac{x}{b}\right) \right ]_0^\infty \qquad \& \qquad \int_0^\infty \mathrm{d}x\frac{\cos (a x)}{x^2 + b^2} = \frac{\pi}{2b}\e^{-ab},
\end{equation}
produces
\begin{equation}
\sigma^2(\tau) = \frac{1}{\SNR^2}\frac{1}{(2\Omega \tau)^2} \frac{\Gamma /4}{\Delta f} [3 - 4 \e^{-\Gamma \tau/2} + \e^{-\Gamma \tau}],
\end{equation}
which reduces to the $\sigma_{y\mathrm{,far}}$ in Section~\ref{allandevdefinition} for short $\tau$.

For long $\tau$ we know (equation~\ref{s:eq:phasenoisenear}) that 
\begin{equation*}
S_{\phi\mathrm{,near}}^\mathrm{th}(\omega) \sim \frac{1}{(\SNR) ^2} \frac{1}{2 \Delta f} \qquad \omega <\!\!< (\Gamma/2)
\end{equation*}
We can integrate this directly in equation~\ref{s:eq:allanvarbasic}
\begin{align}
\sigma^2(\tau) & = \frac{1}{\pi}\left(\frac{2}{\Omega \tau}\right)^2 \int_{0}^{f_H 2 \pi} \mathrm{d}\omega S_{\phi}(\omega) \sin^4\left(\frac{\omega \tau}{2}\right) \nonumber \\
\sigma^2(\tau) & \sim \frac{1}{\pi}\left(\frac{2}{\Omega \tau}\right)^2 \frac{1}{\SNR ^2} \frac{1}{2 \Delta f} \frac{1}{8} \int_{0}^{f_H 2 \pi} [3 - 4\cos(\omega \tau) +\cos(2\omega \tau)] \mathrm{d}\omega \quad \mathrm{for}~\omega <\!\!<\Gamma/2 \nonumber \\
&\approx \frac{3}{2} \frac{1}{\SNR ^2} \left (\frac{1}{\Omega \tau}\right )^2
\end{align}
Here, the $\cos$ terms are assumed to be small:
\begin{equation}
\left. 3\omega\right ]_0^{2\pi f_H} >\!\!> \left. -\frac{4\sin\omega\tau}{\tau}\right]_0^{2\pi f_H} \quad \& \quad \left. \frac{2\sin 2 \omega\tau}{2\tau}\right]_0^{2\pi f_H}.
\end{equation}
In our experiment $f_H = 8/\tau$, so $6\pi f_H$ is $12\times$ larger than the second term, even if $\sin \omega \tau = 1$.

\subsection{Comparison of some literature benchmark equations to the present work}

In this section, for clarity for the community, we rewrite results from well-known works in terms of our definitions (such as $\SNR$) and compare them with the present work.  Cleland's derivation of the Allan deviation is as follows. First,
\begin{equation}
P_\mathrm{c} = \frac{\Omega E_\mathrm{c}}{Q} \quad \& \quad \varepsilon_\mathrm{c} = \frac{2 \pi P_\mathrm{c}}{\Omega \kb T} \qquad \Rightarrow \qquad \varepsilon = 2 \pi 4 \frac{\Delta f}{\Omega} \SNR^2.
\end{equation}
which then leads to
\begin{equation}
\sigma (\tau) = \frac{1}{Q} \sqrt{\frac{\pi}{4\varepsilon_\mathrm{c}\omega \tau}} \qquad \Rightarrow \qquad \sigma(\tau) = \frac{1}{4Q}\frac{1}{\SNR}\frac{1}{(2 \Delta f \tau)^{1/2}}.
\end{equation}

Ekinci uses the frequency stability $\delta \omega/\omega$ for the Allan deviation and gives the following result
\begin{equation}
\left [\frac{\delta \omega}{\omega} \right] \approx \left [ \frac{\kb T}{E_\mathrm{c}}\frac{\Delta f}{\omega_0 Q} \right ]^{1/2} \approx \left [ \frac{1}{\SNR^2}\frac{\Gamma/4}{\Delta f}\frac{\Delta f}{\omega_0 Q}\right]^{1/2}.
\end{equation}
In the above equation, the numerator $\Delta f$ is Ekinci's and the demoninator $\Delta f$ is from the present work and are distinct. They define $\Delta f \equiv 1/(2\pi \tau)$ while we define $\Delta f$ as the explicit instrument bandwidth, i.e. the demodulator effective BW. Also of note, a factor of $2\pi$ in the numerator is dropped during their own approximation during the integration. With our defined $\Delta f$, their $\delta \omega/\omega$ becomes:
\begin{equation}
\left [\frac{\delta \omega}{\omega} \right] \approx \frac{1}{\SNR} \frac{1}{2Q} \frac{1}{(2\pi \Delta f \tau)^{1/2}}.
\end{equation}

Lastly, Gavartin derives the Allan variance as:
\begin{equation}
\sigma^2(\tau)= \left(\frac{1}{\Omega_\mathrm{M} \tau}\right)^2 \frac{\langle x_\mathrm{th}^2  \rangle}{\langle x_\mathrm{d}^2  \rangle} \left[3 - 4\e^{-\Gamma \tau /2} + \e^{-\Gamma \tau}\right].
\end{equation}
Following from this the Allan deviation for long and short $\tau$ becomes, respectively:
\begin{align}
\sigma_\mathrm{long} &= \frac{(3)^{1/2}}{2}\frac{1}{\SNR}\frac{1}{(\Omega Q \Delta f)^{1/2}}\frac{1}{\tau} \\
\sigma_\mathrm{short}&= \frac{1}{2Q}\frac{1}{\SNR}\frac{1}{(\Delta f \tau)^{1/2}}
\end{align}
 
These comparisons between previous work and the current work is shown in Figure~\ref{s:fig:comparison}.

{\renewcommand{\arraystretch}{3}
\begin{figure}\scalebox{0.8}{
\begin{tabular}{l c c c c}
& Cleland & Ekinci & Gavartin & This work \\
\makecell[l]{$S_\phi$} & \makecell{\includegraphics{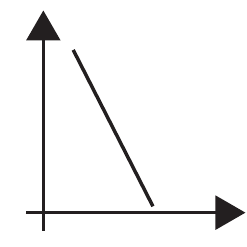}} & \makecell{\includegraphics{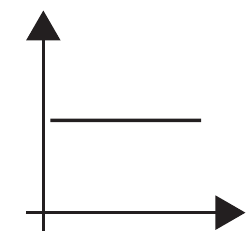}} & \makecell{\includegraphics{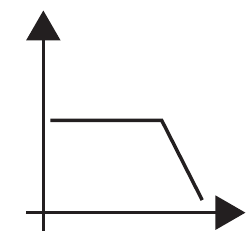}} & \makecell{\includegraphics{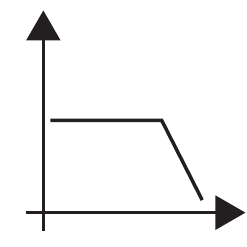}}\\
$H$ & {\setstretch{1.0}\makecell{$\sin^4$\\(implied)}} & 1 & $\sin^4$ & $\sin^4$ \\
{\setstretch{1.0}\makecell{integration\\range}} & $ 0 - \infty$ & $\pm \pi \Delta f$ & \makecell{$0 - \infty (\sigma_\mathrm{short})$\\$0-\infty (\sigma_\mathrm{long})$} & \makecell{$0 - \infty (\sigma_\mathrm{short})$\\$0-2\pi f_H(\sigma_\mathrm{long})$} \\
$\sigma$ short $\tau$ & $\sigma_\mathrm{short} \sim \frac{1}{4Q}\frac{1}{\SNR}\frac{1}{(2 \Delta f \tau)^{1/2}}$ & $\sigma_\mathrm{short} \sim \frac{1}{2Q}\frac{1}{\SNR}\frac{1}{(2 \pi \Delta f \tau)^{1/2}}$ & $\sigma_\mathrm{short} \sim \frac{1}{2Q}\frac{1}{\SNR}\frac{1}{(\Delta f \tau)^{1/2}}$ & $\sigma_\mathrm{short} \sim \frac{1}{4Q}\frac{1}{\SNR}\frac{1}{(\Delta f \tau)^{1/2}}$\\
$\sigma$ long $\tau$ & $\sigma_\mathrm{long} \sim \mathrm{N/A}$ & $\sigma_\mathrm{long} \sim \mathrm{N/A}$ & $\sigma_\mathrm{long} \sim \frac{(3)^{1/2}}{2}\frac{1}{\SNR}\frac{1}{(\Omega Q \Delta f)^{1/2}}\frac{1}{\tau}$ & $\sigma_\mathrm{long} \sim \frac{(3)^{1/2}}{2}\frac{1}{\SNR}\frac{1}{\Omega}\frac{1}{\tau}$ \\
$S_\phi = \xi \frac{S_\mathrm{x}}{\langle x^2 \rangle}$ & $\xi = \frac{1}{2}$ & $\xi = 1$ & $\xi = 1$ & $\xi = \frac{1}{2}$

\end{tabular}}
\caption{\textbf{Comparison of the some literature benchmark equations to the present case.}\label{s:fig:comparison}}
\end{figure}
 }

\subsection{Frequency fluctuation noise}
The frequency stability of the nanomechanical resonator in this work reaches closely to its predicted thermodynamic limit as displayed in Fig. 4. It appears to maintain the thermodynamic limit at short duration $\tau$ (while for longer $\tau$, other noise sources begin to dominate). The presented results are in contrast to the 2016 nature nanotechnology study reported by M. Sansa \textit{et al.}\cite{Sansa2015}. The group reviewed 25 different published works on measured frequency stability of nanomechanical resonators with different designs and sizes and found that none of those devices can attain the experimental stability down to the thermal noise limit by $DR$ formula (equation (1)) in main text). Their study revealed that along with additive thermal noise another source of extra phase noise exists in NEMS class of devices which is parametric and is known as “frequency fluctuation noise”: intrinsic fluctuations in resonance frequency over time that are independent of thermal bath and drive effects. They find this noise to have a flicker behavior following a $f^-1$ power law and giving flat temporal Allan deviation response. If frequency fluctuations noise dominates over additive white noise sources (such as TM noise) frequency stability of a resonator becomes time independent as seen for carbon nanotube\cite{Moser2014}. This extra noise source is independent of the signal to noise ratio. As a consequence, the stability of the device cannot be improved with increasing $\SNR$ , and thus applications of $DR$ formula becomes invalid (see fig.3 in \cite{Sansa2015}). The most obvious sign of frequency fluctuation noise is thus a plateau in the Allan deviation where increasing drive power does not further reduce the deviation. 

In PLL measurements noise suppression occurs due to feedback, and it is hard to distinguish characteristic signature of parametric frequency fluctuation noise. To study the evidence of frequency fluctuation noise, we have used open loop measurements. In open loop, the resonance frequency is locked, and no feedback is applied to close the loop. As a result, there is no deviation of the frequency with time. The collected time trace by lock-in demodulator is a simple quadrature measurement which provides a time stamp for in-phase and quadrature component of amplitude, and the phase at the set frequency. The frequency fluctuations from the locked or set frequency can easily be calculated from the measured phase noise. Since there is no feedback, the measured phase noise is a true characteristic of the device. Such open-loop experiments were performed at different driven amplitudes within the onset of nonlinearity with various lock-in bandwidths. The measured Allan deviations at different pressures are shown in the figure \ref{fig:openADatPs}.

\begin{figure}
    \centering
    \includegraphics[scale=1]{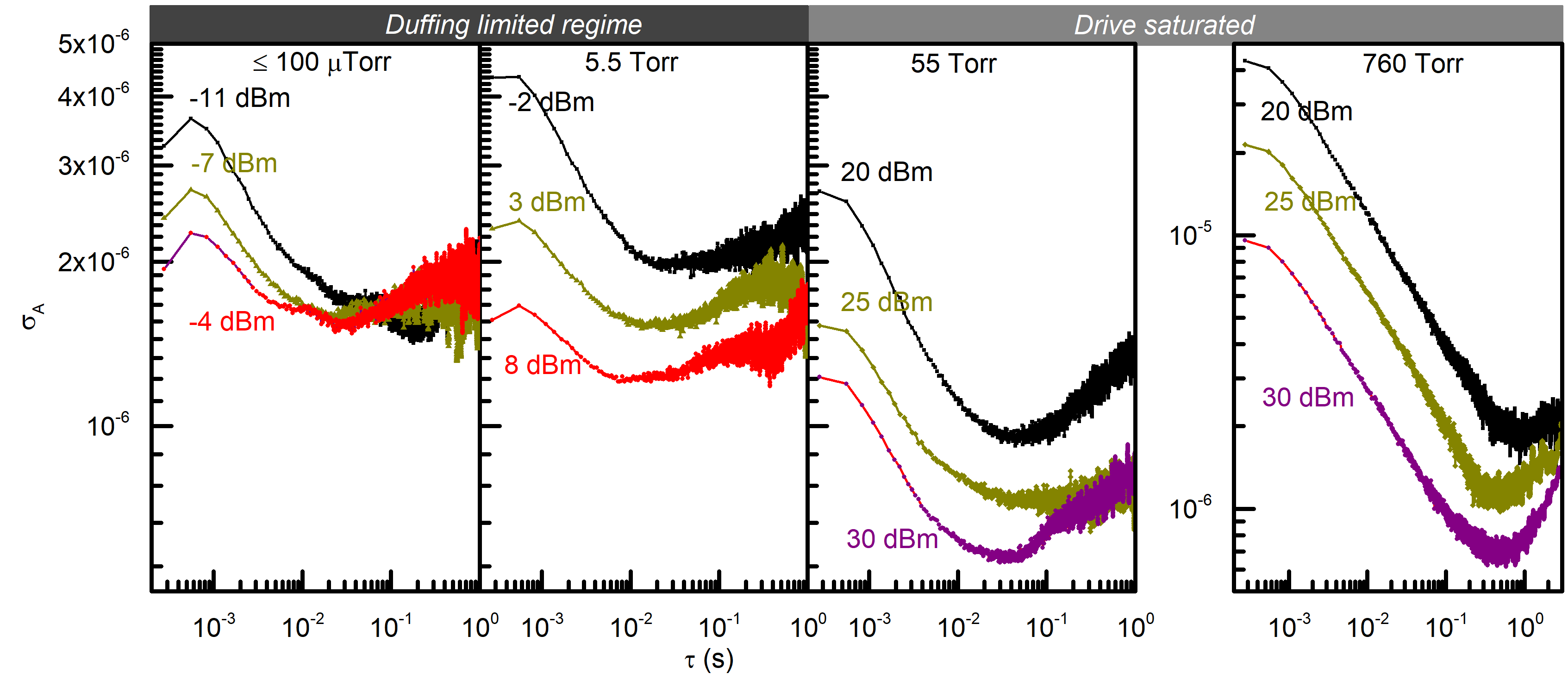}
    \caption{ \textbf{Measured Allan deviations at different pressure regimes and various driving powers with $1$ KHz demodulation bandwidth.} With a sampling frequency of 3600 Hz data is collected for $20$ sec containing $71559$ data points. $20$ s is a relatively large time in comparison with ring downtime ($\sim  \frac{Q}{\omega _{0}}$) of the resonators at each pressure. Falling off in all AD plot below $1$ ms appears because of the roll-off effect of low-pass filtering and data below $1$ ms are not meaningful. Red plots are corresponding Allan deviation measured at the respective critical drive power. At $100$ $ \mu \textup{Torr}$ short-term stability improves with driven amplitudes. At the lowest drive, the -11 dBm data follows $\tau^{-1/2}$ which extended up to 100 ms and is a signature of white noise limited stability. With increasing drive, there is a gradual collapsing of white noise nature ($\tau^{-1/2}$ slope), and the noise floor ($\tau^0$) at all three driven cases are almost similar which is characteristic of frequency fluctuation noise for long measurement time. With increasing pressures (i.e., a decrease in Q) it is evident that effect of frequency fluctuation noise ( collapse of $\tau^{-1/2}$ behavior for shorter averaging time with increasing drive powers) is progressively weakening; the signature of pure additive white noise with additive 1/f noise becomes gradually stronger. The $760$ Torr data show strong evidence of simple additive noise operation which is free from frequency fluctuation noise signature.} 
    \label{fig:openADatPs}
  \end{figure}
  From the figure, it is evident that parametric frequency fluctuation noise for sampling times longer than 20 ms dominates on frequency noise measurements in high $Q$ regime by the silicon NOMS device in this work. The $5.5$ and $55$ Torr data do not show domination by frequency fluctuations, though neither is their behavior fully consistent with additive noise alone.  For 760 Torr, there is no hint of frequency fluctuation noise, and the data are entirely compatible with additive noise sources.
 \begin{figure}[h]
\centering
\includegraphics[scale=1]{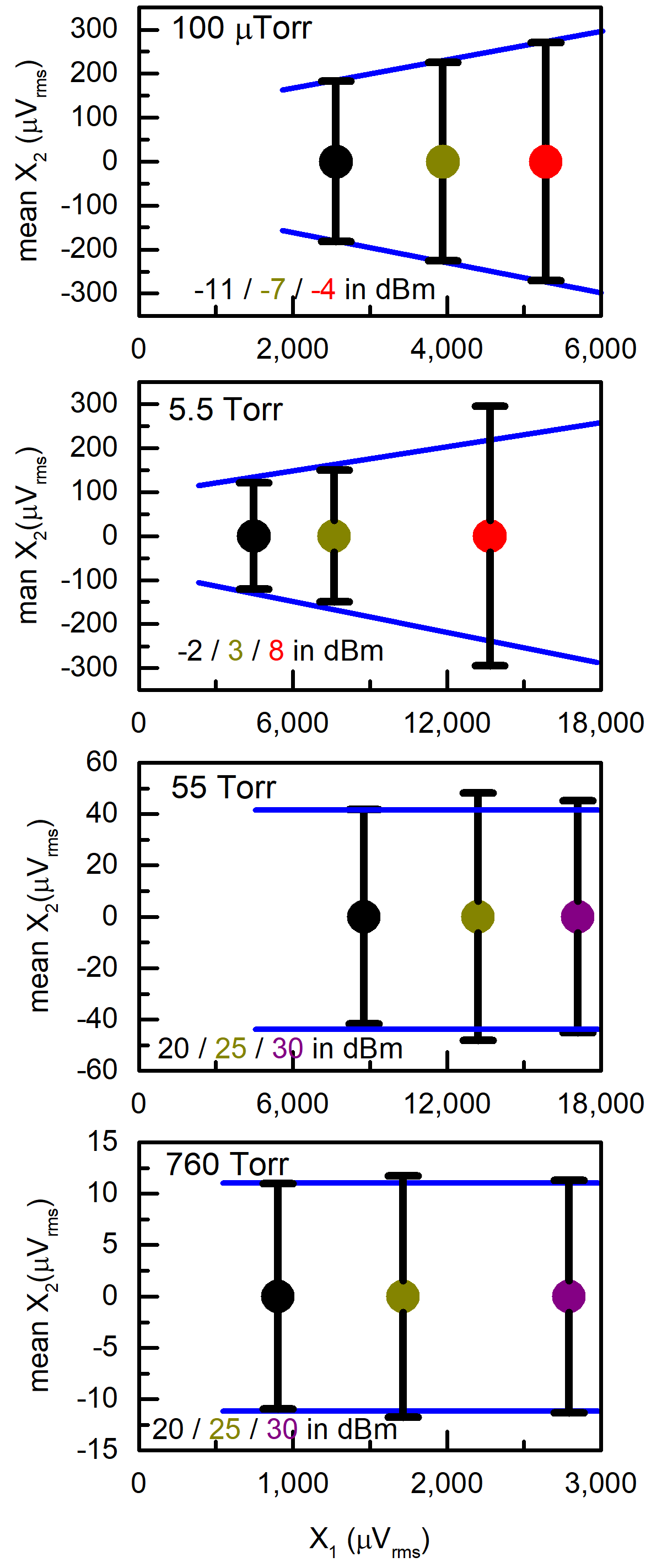}
\includegraphics[scale=0.95]{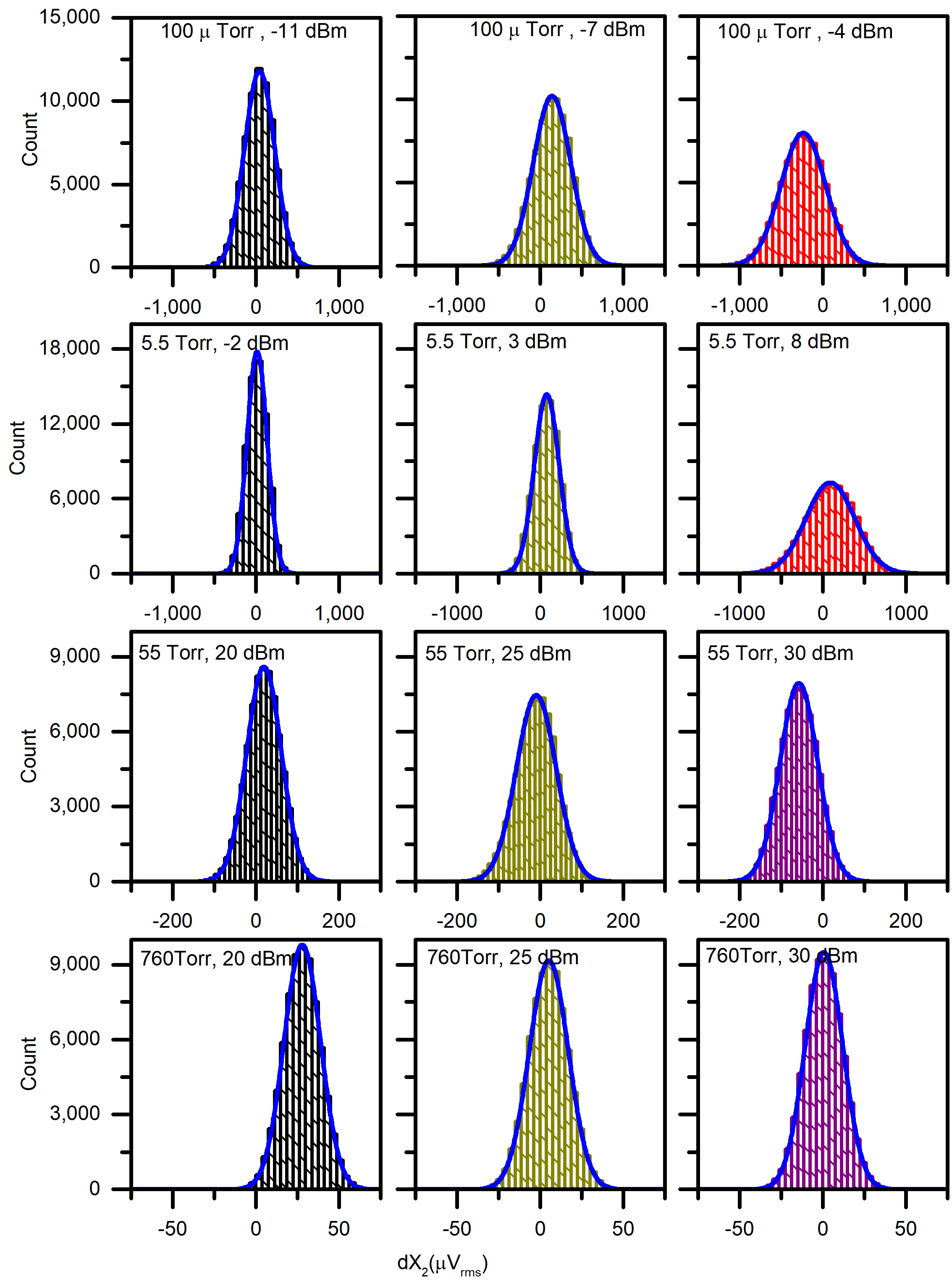}
\caption[Figure S2 short description]{\textbf{Left:  Quadrature representation of the same data used in figure (\ref{fig:openADatPs}) to calculate Allan deviations.} The in-phase-quadrature ($X_{1}$)) at each data set is rotated in order to make the mean phase zero so that data can be centered at zero and at the same time, the amplitude of oscillations (in $\mu \textup{V}$) can be plotted along the horizontal axis. Consequently, phase information ($X_{2}$)  in $\mu \textup{V}$)) can be set along the vertical axis. Mean phase for each dataset is forced to zero to show the variation of phase quadrature noise by the standard deviation (black bars) of$X_{2}$) data with respect to driven amplitude, $X_{1}$). Blue lines are guides to the eye. For lower pressure and lower damping (higher $Q$), phase quadrature noise increases with driven amplitude; this is in contrast to 760 Torr data at higher damping where phase quadrature noise is constant with driven amplitude.
\textbf{Right:  Histograms of the phase quadrature values.} The blue line at each histogram is a normal fit for each set. Widening and shortening of the histograms at higher drives confirm the presence of excess frequency fluctuation noise for higher $Q$. Width and height are constant for 760 Torr showing that frequency fluctuation is negligible at atmospheric pressure.}
 \label{fig:openquad}
\end{figure}

To confirm these findings, we plot both standard deviations and histograms of the phase quadrature as a function of drive power in figure \ref{fig:openquad}. These data are for the full 20-second datasets, so they incorporate behavior from all averaging times $\tau$. Vacuum data show growth in the standard deviation of the phase quadrature ($X_{2}$) fluctuations with drive power, another signature of frequency fluctuation being the primary noise source. In the atmospheric pressure case, phase quadrature deviations remain the same with respect to driven amplitudes. Intermediate pressures show some effect of a noise source that is not diminished with drive power (such as frequency fluctuations). In all cases, the phase angle lines do not converge at zero drive, so frequency fluctuation noise is never the sole noise source in any case. The histograms support similar conclusions. With increasing drive power, the histograms shorten and widen for the two lower pressures, and remain constant for atmospheric pressure. $55$ Torr histograms reflect almost similar behavior as by $760$ Torr data, but Allan deviation plots at different driving power do not exactly proportionally decrease with driving amplitude, which is an indication of excess noise over thermal noise at this pressure.

From the Allan deviation data, we can infer that the frequency fluctuation noise is only kicking in for longer averaging times. This would be consistent with the noise source being temperature fluctuations of the DCB, especially considering our very large temperature coefficients with frequency. As such, this effect might be partially mitigated at atmospheric pressure by the much larger heat transfer coefficient with the surrounding air. We also note that frequency fluctuation noise should translate into phase noise in proportion to $Q$ (see Ref.\citenum{Fong2012}), which gives another reason for the effect being indiscernible at atmospheric pressure. In any case, the effect of frequency fluctuations is negligible for 2 ms averaging time, as evidenced by Figure 4 in main text.

\clearpage

\putbib[DR_bibliography]
\clearpage
\end{bibunit}

\end{supplementary}


\end{document}